\documentclass[usenatbib]{mnras}
\usepackage[T1]{fontenc}
\usepackage{ae,aecompl}
\usepackage{amsmath}
\usepackage{amssymb}
\usepackage{mathtools}
\usepackage{bm}
\usepackage{cleveref}
\usepackage{graphicx}
\usepackage{caption}
\usepackage{subcaption}
\usepackage{float}
\usepackage{txfonts}
\usepackage{hyperref}
\usepackage{multirow}
\usepackage{hhline}
\title[Extra Galactic  GCs - BH population in MW \& M31]
{MOCCA-Survey Database: Extra Galactic Globular Clusters. III.  The population of black holes in Milky Way and Andromeda - like galaxies}
\author[A. Leveque et al.]{
A.~Leveque$^{1}$\thanks{E-mail:agostino@camk.edu.pl},
M.~Giersz$^{1}$,
Abbas Askar$^{2}$,
M. Arca-Sedda$^{3}$
\& A. Olejak$^{1}$
\\
$^{1}$ Nicolaus Copernicus Astronomical Center, Polish Academy of Sciences, ul. Bartycka 18, PL-00-716 Warsaw, Poland\\
$^{2}$ Observatory, Department of Astronomy, and Theoretical Physics, Lund University, Box 43, SE-221 00 Lund, Sweden\\
$^{3}$ Department of physics and astronomy "G.Galilei", University of Padua, Vicolo dell'Osservatorio 3, I-35122, Padua, Italy \\
}
\graphicspath{{Figures/}}

\begin{document} 

   \date{}
   \pagerange{\pageref{firstpage}--\pageref{lastpage}} \pubyear{2022}
   \maketitle

 
\begin{abstract}
This work investigates the black hole (BH) population of globular clusters (GCs) in Milky Way- and Andromeda- like galaxies. We combine the population synthesis code MASinGa and the MOCCA-Survey Database I to infer the properties of GCs harbouring  a stellar-mass BH subsystem (BHS), an intermediate-mass BH (IMBH), or neither of those. We find that the typical number of GCs with a BHS, an IMBH, or none become comparable in the galactic outskirts, whilst the inner galactic regions are dominated by GCs without a significant dark component. We retrieve the properties of binary BHs (BBHs) that have either merged in the last 3 Gyr or survived in their parent cluster until present-day. We find that around 80\% of the merging BBHs form due to dynamical interactions while the remaining originate from evolution of primordial binaries. The inferred merger rate for both in-cluster and ejected mergers is  $1.0-23\,\,\rm{yr^{-1}\,Gpc^{-3}}$ in the local Universe, depending on the adopted assumptions. We find around 100-240 BBHs survive in GCs until present-day and are mostly concentrated in the inner few kpc of the galaxy. When compared with the field, GCs are at least two times more efficient in the formation of BHs and binaries containing at least one BH. Around $1,000-3,000$ single BHs and $100-200$ BBHs are transported into the galactic nucleus from infalling clusters over a time span of 12 Gyr. We estimate that the number of BHs and BBHs lurking in the star cluster to be about $1.4-2.2\times10^4$ and $700-1,100$ respectively.
\end{abstract}


%

\section{Introduction}
Numerous observational studies have found black holes (BHs) and accreting BHs candidates in Galactic and extragalactic globular clusters (GCs) \citep{Maccarone2007,Barnard2009,Roberts2012,MillerJones2015,Minniti2015,Bahramian2017,Dage2018}. Radial velocity measurements of a binary system in NGC 3201 provide the strongest proof that a BH exists in a Galactic GC \citep{Giesers2018,Giesers2019}. Additionally, both electromagnetic emission and dynamical mass measurements from kinematic observations of extragalactic GCs indicate the presence of a significant fraction of unseen mass, possibly stellar-mass BHs in binary systems and intermediate-mass BH (IMBH) \citep{Taylor2015,Dumont2022}. Similarly, recent studies have also looked for indicators of the presence of BHs in GCs using numerical simulations of GC models containing sizable populations of BHs \citep{Morscher2015,ArcaSedda2016,Askar2018,ArcaSedda2019,Weatherford2019}.

Generally, the most massive GCs should form up to several thousands of BHs in the first few Myr of cluster evolution. The natal kicks that these BHs experience upon birth and the cluster's escape velocity have a substantial impact on how many of these BHs can be kept in the GCs, and it is crucially affected by the uncertain physics of stellar collapse \citep{Belczynski2002,Belczynski2010,Fryer2012,Repetto2017,OShaughnessy2017}. Retained BHs would segregate rapidly, populating the central regions of their host GC \citep{Portegies2000,Portegies2002,Fregeau2004,Freitag2006,ArcaSedda2016}. The dynamics in the cluster central region is thus dominated by stellar-mass BHs, and can lead to different outcomes. One possibility is that dynamical interactions among the most massive BHs lead to their ejection, freeing the cluster centre from their dark content. Another possibility is that, despite dynamics, BHs form a subsystem dominating the innermost cluster region, or they merge among themselves or with other stars to build-up an IMBH. IMBHs have masses in the range $10^2 - 10^5 \,M_\odot$ and they are considered as the link between the stellar-mass and supermassive BHs \citep{Barack2019}. Recent numerical studies 
\citep{BreenandHeggie2013a,BreenandHeggie2013b,Heggie2014,Webb2018,Banerjee2018,Kremer2018,Kremer2019} showed that the BHS is not entirely decoupled from the rest of the GC, and that the energy demands of the host GC would control the evolution of its BHS \citep{BreenandHeggie2013a,BreenandHeggie2013b}. The presence of a massive BHS can be responsible for the cluster dissolution,
due to the interplay of the strong energy produced from the BHS and tidal stripping \citep{Giersz2019}. 

 High stellar densities are required to form massive BHs such as IMBH, with a possible scenario of IMBH formation being repeated collisions in the central regions of GCs \citep{PortegiesZwart2002,PortegiesZwart2004,Portegies2007,Giersz2015,Mapelli2016}. Indeed, IMBH can be form through multiple stellar mergers in binary system \citep[see also \citealt{Maliszewski2022}]{DiCarlo2021,Gonzalez2021,ArcaSedda2021,Rizzuto2021,Rizzuto2022}. So far, there is still no conclusive evidence of IMBH presence in Galactic GCs, even though they are considered to potentially host an IMBH \citep{Bash2008,Maccarone2008,Lutzgendorf2013,Lanzoni2013,Kamann2014,Askar2017,ArcaSedda2019,Hong2020,ArcaSedda2020}.

Multiple three-body interactions can cause the formation of BH-BH binaries (BBHs), which serve as a power supply for the cluster core. The continuous interactions between the BBHs and the other objects in the GC, would harden the binaries, until they would be ejected from the cluster core or be merged, releasing gravitational waves \citep{Portegies2000,Banerjee2010,Downing2010,Wang2016,Askar2017}. Similarly, stars that interact with retained BHs are forced into wider orbits, causing the GC to expand and this can postpone core-collapse \citep{Merritt2004,Mackey2008,Gieles2010,Wang2016,Kremer2019}. The presence of a BHS or of a IMBH in the central region of a GC would importantly shape the structure of the host GC \citep{Mackey2007,Zocchi2015,ArcaSedda2018,Baumgardt2020}.

In this work we extend the study of the GC populations for the MW and M31 exploited in our companion paper \citep[herefter Paper II]{Leveque2022b}. In previous papers in this series, we set up the machinery that would be used to populate external galaxies with their GC populations by combining the results from the MOCCA-Survey Database I with the MASinGa semi-analytic tool. In this work we would like to test for the first time our machinery against the GCs properties and their BH content simulated in our models for both MW and M31 populations. In particular, we compare the orbital properties of the MW GC population with the observed properties from the Bajkova catalogue \citep{Bajkova2021}. Also, we aim to constrain the spatial distribution in the galactic halo of different GCs properties for different GC dynamical states comparing our results with previous studies \citep{Lutzgendorf2013,Askar2018,ArcaSedda2019,Weatherford2019}. Then, we intend to determine the properties of the BBH mergers reported in our simulations and the inferred BBH merger rate \citep{Banerjee2022,Mapelli2022}. Finally, we show the properties of BBHs present at 12 Gyr that could potentially be observed, and the number of BH and BH binaries that have been transported to the nuclear star cluster (NSC) by infalling star clusters. In Appendix \ref{Appendix} we present the statistical tests of the studied populations.

\section{Method} \label{sec:Method}
In this section, we will summarize the most important ingredients of our machinery. More details about all the physical assumptions are properly described in section 2 of Paper II \citep{Leveque2022b}.

The MASinGa  (Modelling Astrophysical Systems In GAlaxies) program has been used to model the GC populations \citep{ArcaSedda2014-2,Belczynski2018,Leveque2022b,ArcaSedda2022}. For each GC in the population, MASinGa simulates the orbital evolution while taking into consideration the galactic tidal field and shocks, which contribute to the cluster disintegration, dynamical friction, which pulls the cluster toward the galactic center, and internal relaxation, which controls the cluster mass loss and expansion/contraction. 

The number of GCs and distributions of GC masses and galactocentric positions generated by MASinGa have been used to choose appropriate models from the MOCCA-Survey Database I  \citep{Askar2017} to reproduce the initial distributions (the details are provided in Sec. 2.3 in Paper II).
However, few steps have been taken in order to determine the galactocentric position of the MOCCA models in the gravitational potential of the studied galaxy. Indeed, the simple point-mass approximation for the Galactic potential was used to evolve the MOCCA models, with the central galaxy mass being contained inside the GC's orbital radius. The initial Galactocentric distance of a model in MOCCA-Survey Database I is defined by its initial mass and its tidal radius. Only a finite set of such values were considered \cite[see Table 1 in][]{Askar2017}, with no specific initial density profile being taken into account while modeling the MOCCA-Survey Database I models. 
A circular orbit at Galactocentric distances between 1 and 50 kpc were assumed, and the GC's rotation velocity was set to to $220$ $\rm{km\,\,s^{-1}}$ for the whole range of galactocentric distances. The correct galactocentric distance for a circular orbit in an external galaxy, for a given tidal radius $r_{tidal}$ and GCs mass $M_{GC}$ can be determined knowing the tidal radius for the MOCCA model and the density distribution of the simulated galaxy.
From the results presented in \cite{Cai2016}, it is possible to determine the  galactocentric radius of a circular orbit on which a GC will experience an equivalent mass loss if it were on an eccentric orbit. In particular, the apocenter distance $R_{apo}$ for the eccentric orbit can be determined as $R_{apo} = R_C\cdot(1 - 0.71 \cdot E_{GC})^{-5/3}$ \citep{Cai2016}, with $R_C$ the galactocentric distance for a circular orbit, and $E_{GC}$ the initial orbital eccentricity, chosen from a thermal distribution. The initial galactocentric distance for a circular orbit  $R_C$ is known for the MOCCA models. Finally, the initial galactocentric position has been selected within the orbit apsis. For each MOCCA model, a total of 900 representations of each MOCCA-Survey Database I model were generated, with different initial orbital eccentricity and galactocentric distances - the MOCCA-Library. In this way, the same model could be populated with different orbital parameters and in different galactocentric distance regions. Each simulated cluster from the MOCCA-Survey Databse I has a different internal dynamical evolution (such as mass loss, half-mass radius, etc.). Consequently, the dynamical evolution of MOCCA-Library models that represent different MOCCA models are diverse. Hence, models from the MOCCA-Library are defined as unique when they represent different MOCCA models (more details are provided in Sec. 2.3.1 in Paper II \citep{Leveque2022b}).

Only MOCCA models that survived their internal dynamical evolution up to 12 Gyr were selected to represent the MASinGa GC models. Indeed, the initial galactocentric position for each MOCCA model was chosen within the same initial galactocentric radius bin of the representative MASinGa model. Instead, the initial mass was chosen randomly from the GC initial mass function (GCIMF) cumulative distribution (a power law $dN/dm = b \cdot m^{-\alpha}$ with a slope of $\alpha = 2$ function has been used as GCIMF), with lower and upper limits of $2\times10^5 \,\,M_\odot$  and $1.1\times10^6 \,\,M_\odot$ respectively. This mass range was set in order to reproduce the expected initial mass range for MW and M31, being of $\sim 10^5 - 10^7$. Indeed, the observed masses for the survived GCs located within 17 kpc from the galactic center range between $10^4 - 2\times10^6 \,\,M_\odot$ and $5\times10^4 - 3\times10^6 \,\,M_\odot$, for MW and M31 respectively\footnote{In this calculation, we did not take in consideration GCs with masses at 12 Gyr smaller than $10^4 \,\,M_\odot$, as they would be closed to the cluster dissolution, and they would be hard to be observed in external galaxies.}. According to \cite{Webb2015}, the initial GC mass was $\sim 5$ times greater than the actual observed values (more details are provided in section 3.3 of Paper II). On the other hand, the maximum mass in the GCIMF was chosen according to the maximum initial mass in the MOCCA models, being of $1.1\times10^6 \,\,M_\odot$.\footnote{For an initial power-law GCIMF between $M_{low} = 10^3-10^4\,M_\odot$ and $M_{up}=10^7\,M_\odot$, we would expect a total number of low-mass clusters ($M < 10^5\,M_\odot$) being around 1500-200 for $M_{low} = 10^3\,M_\odot$ and $10^4\,M_\odot$ respectively, and a total mass of the GCs population around $2-3\times10^8\,M_\odot$.} The GCIMF cumulative distribution has been normalized to the initial MASinGa models mass distribution. Each model representation was successively removed from the MOCCA-Library, in order to avoid multiple instances of the same model to populate each mass and galactocentric position bin. Our procedure would guarantee that the total number of GCs in the population, the galaxy density distribution, and the GC IMF distribution for the MOCCA population would reproduce the initial conditions in the MASinGa population. Also, this procedure guaranteed an 85\% minimum coverage of not repeated unique MOCCA models for both MW and M31 populations.  Finally, most of the selected MOCCA models has an initial pericenter distance within 4 kpc from the galactocentric centre, and a initial thermal distribution for the eccentricity. The models that has been delivered to the NSC have small pericenter distances ($< 0.5$ kpc) and really eccentric orbits ($> 0.8$). The galaxy density profile have been modelled with a \cite{Dehnen1993} density profile family, of the form:
\begin{equation*}
    \rho_G(r) = \frac{(3-\gamma)M_g}{4\pi r_g^3} \left( \frac{r}{r_g} \right)^{-\gamma} \left( 1 + \frac{r}{r_g}\right)^{\gamma-4},
\end{equation*}
where $M_g$ is the galaxy total mass in $M_\odot$, $r_g$ the galaxy length scale in kpc and $\gamma$ the density profile slope. The Dehnen models parameters' values have been determined fitting the observational rotational curve (data from \citep{Eilers2019}  and \citep{Chemin2009}, for MW and M31 respectively) to the Dehnen rotational curve. The initial number of GCs in our models has been set by the total GC mass (defined as a fraction of the total galaxy mass) and the GCIMF. Even though the initial total number of GCs found in our simulations is much smaller compared to the observed total number of GC in both MW and M31, we found a similar number of initial GCs  in our simulations and the observed one when we limit to GCs located within 17 kpc from the galactic center.  Also, we found that the number of surviving GCs in our models is similar to the observed number of GCs within the selected mass and distance range for both MW and M31. The initial conditions used in our simulations to reproduce the MW and M31 population are summarized in Table \ref{Table:init-conditions}.

\begin{table}
    \centering
    \begin{tabular}{ ccc}
    \hline
        Parameter & MW & M31 \\
        \hline
        Galaxy density profile & \cite{Dehnen1993} & \cite{Dehnen1993}\\
        $M_g [M_\odot]$ &$3.18\times 10^{11}$ & $5.75\times 10^{11}$ \\
        $r_g$ [kpc]& 5.12 & 5.8\\
        $\gamma$ &0.54 & 0.1 \\
        GCIMF function & powerlaw & powerlaw\\
        GCIMF slope & 2 & 2\\
        GCIMF $M_{min}\,\, [M_\odot]$ & $2\times10^5 $ & $2\times10^5$  \\
        GCIMF $M_{max}\,\, [M_\odot]$ & $1.1\times10^6 $& $1.1\times10^6$ \\
        $N_{ini}$ & 132 & 245\\
        $M_{GC,ini}\,\, [M_\odot]$ & $6.3\times10^7 $ & $10^8$   \\
    \hline
    \end{tabular}
    \caption{Initial conditions for the MW and M31 GC population simulated in MASinGa. $N_{ini}$ and $M_{GC,ini}$ represent the initial total number and initial total mass of the GC population, respectively. } 
    \label{Table:init-conditions}
\end{table}

Successively, the  MOCCA models selected with the prescription described above have been used to follow the internal and external dynamics evolution of the GC population. This allowed not only to follow the actual evolution for GC's  mass, and half-mass radius but also to follow the evolution of the compact objects present in the system, together with the compact object binary evolution and their survival.
On the other hand, the formulae adopted in MASinGa have been used to determine the galactocentric distance and eccentricity evolution.

It is important to underline that the stellar evolution prescription adopted in MOCCA models is outdated. The BH masses prescription used in MOCCA models follows the \cite{Belczynski2002} mass fallback formulae. In particular, proposed rapid and delayed supernova mechanisms \citep{Fryer2012} were not implemented in stellar evolution prescriptions used in MOCCA-Survey Database I models. For this reason, the final BH masses are  smaller compared to the observed values from GW detections \citep{Abbott2021,Abbott2022} and updated stellar and binary evolution prescriptions for BH progenitors \citep{Kamlah2021}. Also, the sub-sample of MOCCA models used to populate the studied galaxies have metallcities $Z$ of 0.02, 0.006, 0.005, 0.001, and 0.0002. ($Z_\odot = 0.02$).

As mentioned in Paper II and in \cite{Madrid2017}, the MOCCA results were able to recreate the N-body simulations for galactocentric distances down to a few kpc. For this reason, the region between $2$ and $17$ kpc has been the focus of the post-processing investigation and statistical analysis of GC populations' properties in Paper II. In a similar way, we also restricted our analysis in this work to GCs that were located in the same galactocentric zone. 

\section{Results} \label{sec:Results}
For both MW and M31, 100 galaxy models were created, all formed 12 Gyr ago and evolved until the present day as in the prescription described in Paper II. To reduce statistical fluctuations and have a more robust statistical representation of the models, the average values obtained from the galaxy models and their GC population have been considered. The repeating of the same unique model might distort the structural GC parameter distribution, biassing the simulated distribution toward the attributes of the unique models that were randomly picked the most. To prevent such bias, when estimating the radial distribution of each property, only one unique model inside each radial bin was examined. The average value of each population's measurements, as well as the standard deviation, have been calculated for each attribute.

While in Paper II we discussed and studied the global properties of the GC population reproduced by our machinery (such as mass spatial distribution, half-light radius distributions, etc.), in this work we focus on studying the properties of the simulated GC population and their BH content. 

\subsection{Orbital properties in MW GC population}
The orbital properties obtained in the MW GC population have been compared to the observed data from the Bajkova catalogue \citep{Bajkova2021}. The Bajkova catalogue contains the orbital properties of 152 GCs in the MW, determined using the Gaia DR2 proper motions and the data from the \cite{Vasiliev2019} and \cite{Massari2019} catalogs. The orbital properties have been determined considering an axisymmetric Galactic potential based on the Navarro-Frenk-White dark halo \citep{Navarro1997}.

\begin{figure}
    \centering
        \includegraphics[width=\linewidth]{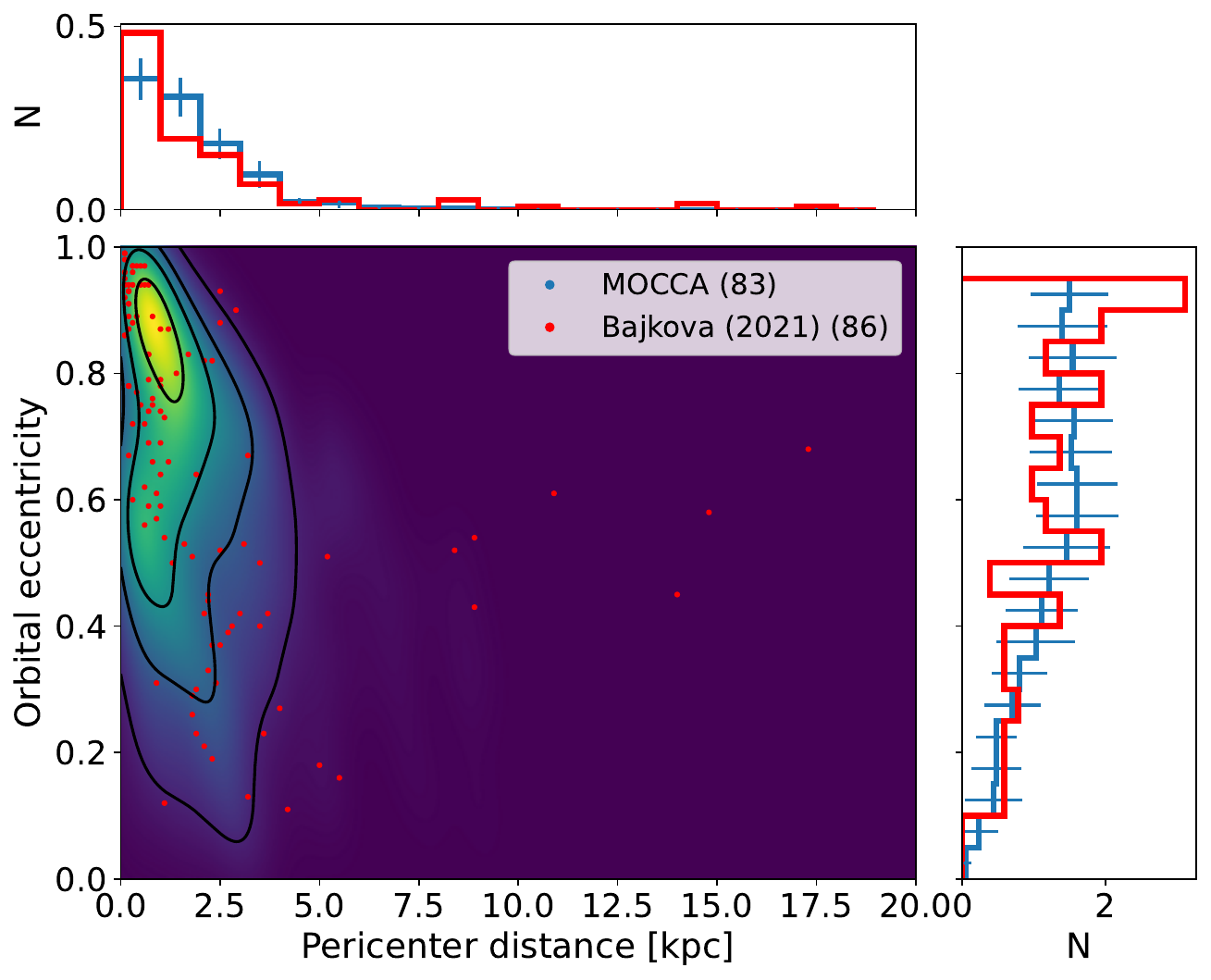}
    \caption{Density map for the orbital eccentricity and pericenter distance for the MOCCA population for MW. The contours include the 80, 50, 30 and 10\% levels of population. The Bajkova catalogue is reported in red. On the side, the normalized histogram showing the distributions of each population is reported (the area beneath the histogram has been set to 1), with the error bars showing the standard deviations for the simulated models. In brackets the mean number of MW GCs simulated and the observed population number have been reported, for MOCCA and Bajkova catalogue respectively.}
    \label{Fig:densityMapBajkowaPerEccMW}
\end{figure}

\begin{figure}
    \centering
        \includegraphics[width=\linewidth]{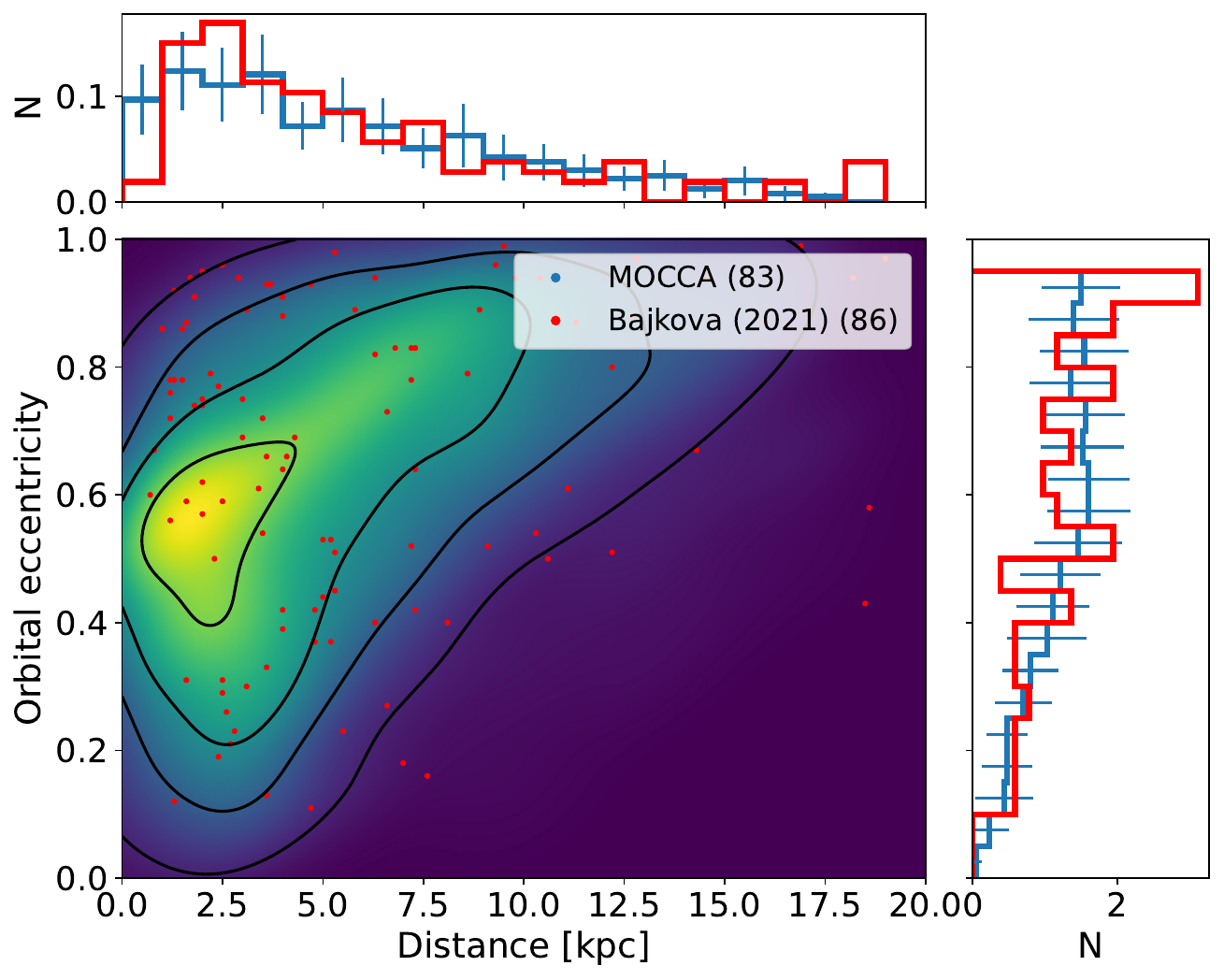}
    \caption{Density map for the orbital eccentricity and galactocentric distance for the MOCCA population for MW. The contours include the 80, 50, 30 and 10\% levels of population. The Bajkova catalogue is reported in red. On the side, the normalized histogram showing the distributions of each population is reported (the area beneath the histogram has been set to 1), with the error bars showing the standard deviations for the simulated models. In brackets the mean number of MW GCs simulated and the observed population number have been reported, for MOCCA and Bajkova catalogue respectively.}
    \label{Fig:densityMapBajkowaEccRadDistMW}
\end{figure}

\begin{figure}
    \centering
        \includegraphics[width=\linewidth]{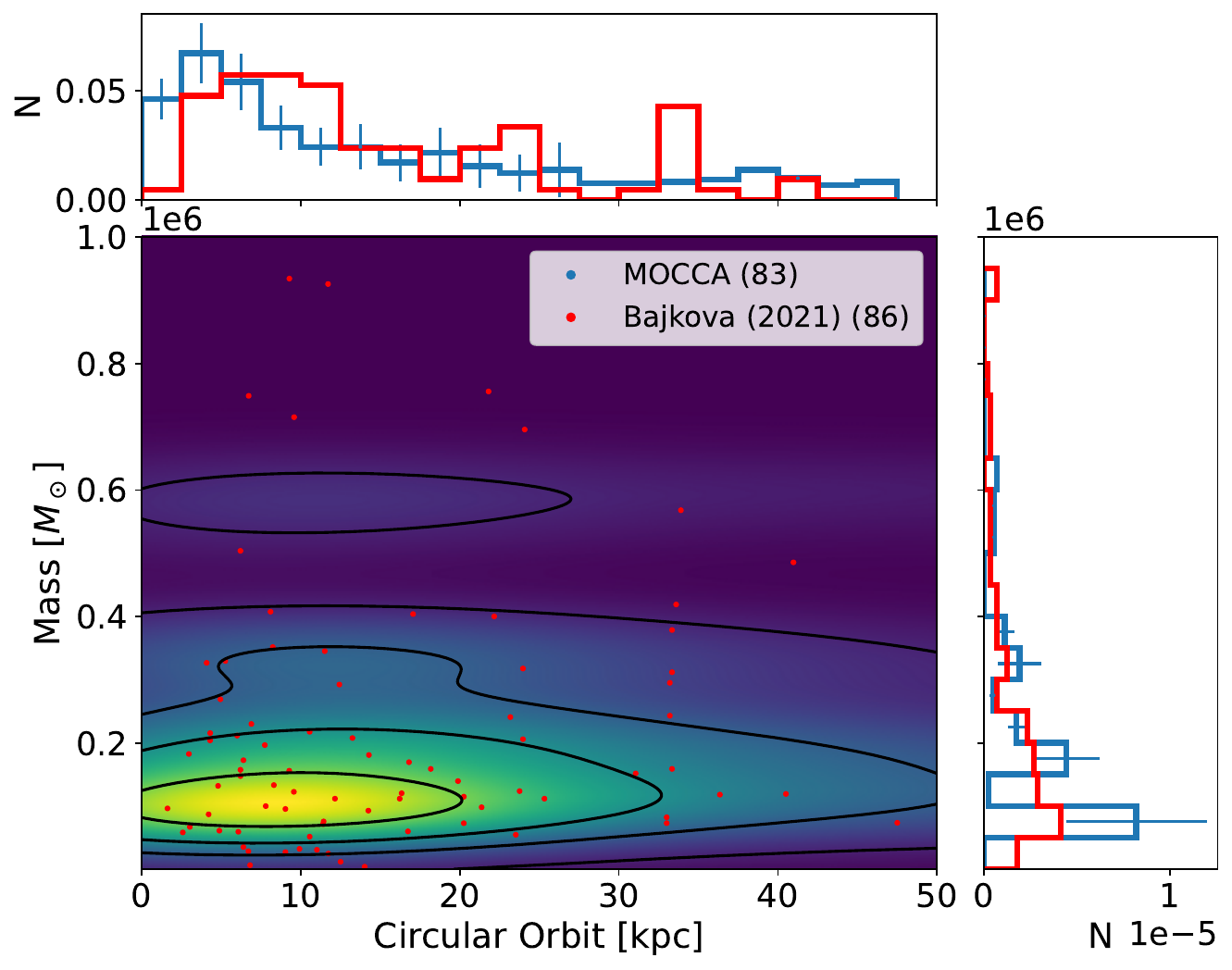}
    \caption{Density map for the circular orbit and the mass for the MOCCA population for MW. The contours include the 80, 50, 30 and 10\% levels of population. The Bajkova catalogue is reported in red. On the side, the normalized histogram showing the distributions of each population is reported (the area beneath the histogram has been set to 1), with the error bars showing the standard deviations for the simulated models. In brackets the mean number of MW GCs simulated and the observed population number have been reported, for MOCCA and Bajkova catalogue respectively.}
    \label{Fig:densityMapBajkowaRcMassDistMW}
\end{figure}

In Fig. \ref{Fig:densityMapBajkowaPerEccMW} we show the density map for the orbital eccentricity and pericenter distance for the population of MOCCA models in the case of the MW. The colour map shows the density map for our models, and the contours include the 80, 50, 30 and 10\% levels of population. In red, we reported the properties retrieved from the Bajkova catalogue \citep{Bajkova2021}. The regions containing most of the populations are presented with brighter colours. On the side, the histogram showing the distribution for each population is also reported, with the area below the histogram being set to 1. Similarly, in Fig. \ref{Fig:densityMapBajkowaEccRadDistMW} and  \ref{Fig:densityMapBajkowaRcMassDistMW} the density map for the projected galactocentric distance versus the eccentricity and for the circular orbit versus the GC mass are reported. The circular orbit of the observed GC has been determined using the results in \cite{Cai2016}. The comparison shows that our sampled models are in reasonable agreement with the observed orbital properties of MW clusters. However, our models have smaller mean values compared to the observed ones (the statistical test results are reported in Appendix  \ref{Appendix}). In our machinery, the GCs have been populated around a galaxy at first on a circular orbit, and successively modelled in elliptical orbits using the results from \cite{Cai2016}. The circular orbit comparison, and more in general the kinematical findings shown so far provide additional confidence  that our machinery can recreate real kinematical properties of MW GCs, and it may also be used to populate external galaxies.

\subsection{Dynamical models}

The presence of an IMBH, or of a BHS, or neither of them, has a significant impact on the dynamical history and characteristics of the GCs. Following the division described in Paper I \citep{Leveque2021}, we divided our chosen sample into three dynamical sub-samples namely:
\begin{itemize}
    \item If there is an IMBH (BH with mass greater than 500 $M_\odot$), the system has been classified as a \textbf{IMBH model};
    \item if the number of BH ($N_{BH}$) present in the system is $\ge 50$, it has been classified as \textbf{BHS model}; if $20 < N_{BH} < 50$, we checked if the system is not experiencing the core collapse: if the system is in balanced evolution \citep{BreenandHeggie2013a,BreenandHeggie2013b}, it has been also classified as \textbf{BHS model};
    \item  a model that is not categorized as an IMBH nor as a BHS has been classified as a \textbf{Standard model}.
\end{itemize}

\begin{figure}
    \centering
    \begin{subfigure}{0.5\textwidth}
       \centering
        \includegraphics[width=\linewidth]{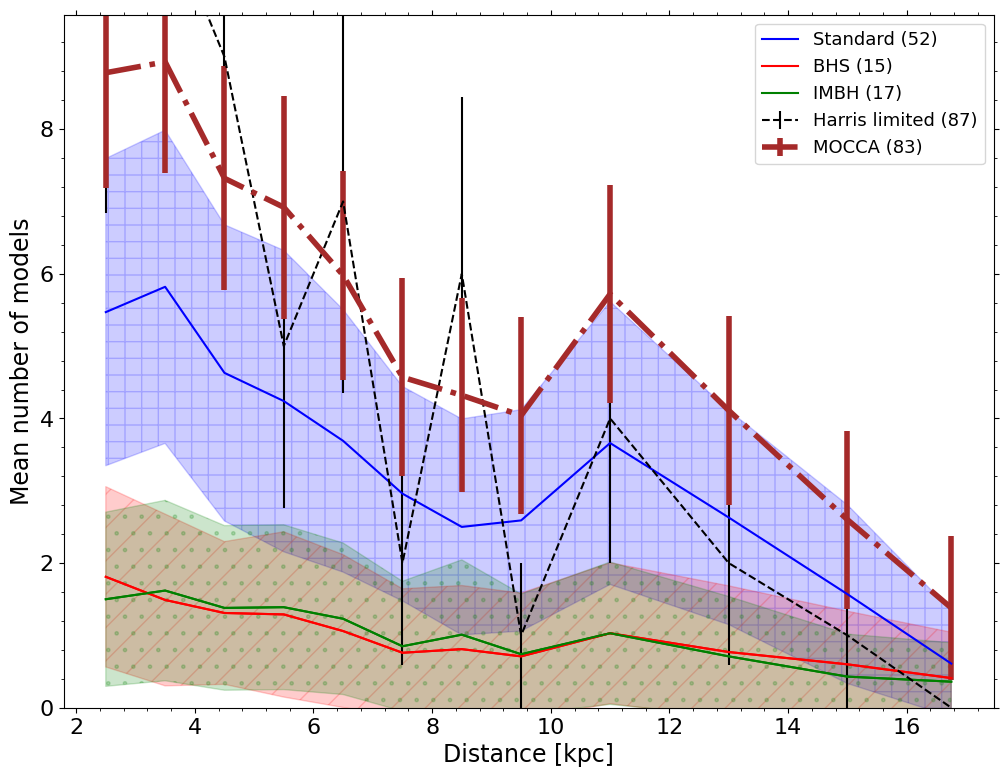}
    \end{subfigure}
    \begin{subfigure}{0.5\textwidth}
       \centering
        \includegraphics[width=\linewidth]{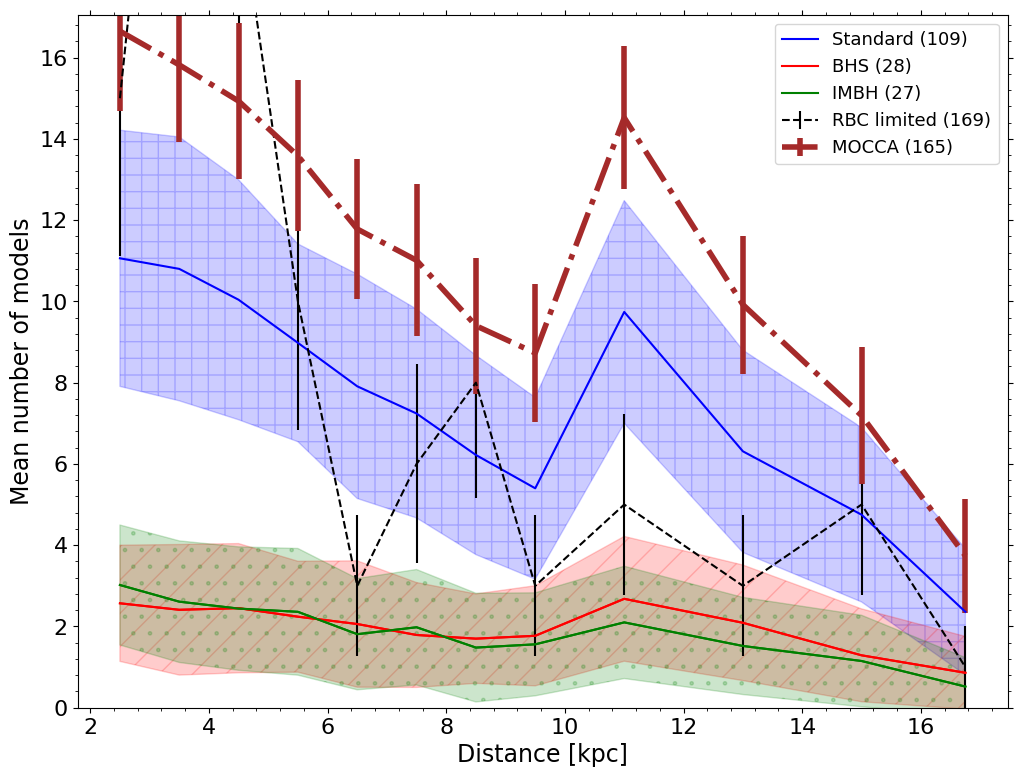}
    \end{subfigure}
    \caption{Spatial distribution for Standard (blue), BHS (red) and IMBH (green) models for MW (top) and M31 (bottom) for the MOCCA,  respectively. The shaded regions represent the standard deviation for the simulated GC populations. The squared region, the oblique lines and the dots show the standard deviation for Standard, BHS and IMBH model respectively. The mean number of GCs for each dynamical model are reported in brackets. In black dashed lines we report the spatial distributions for the observed populations, while in thick brown dot-dashed line we report the spatial distribution for the entire simulated population.}
    \label{Fig:dynamicalModelNDistribution}
\end{figure}

\begin{figure}
    \centering
    \begin{subfigure}{0.5\textwidth}
       \centering
        \includegraphics[width=\linewidth]{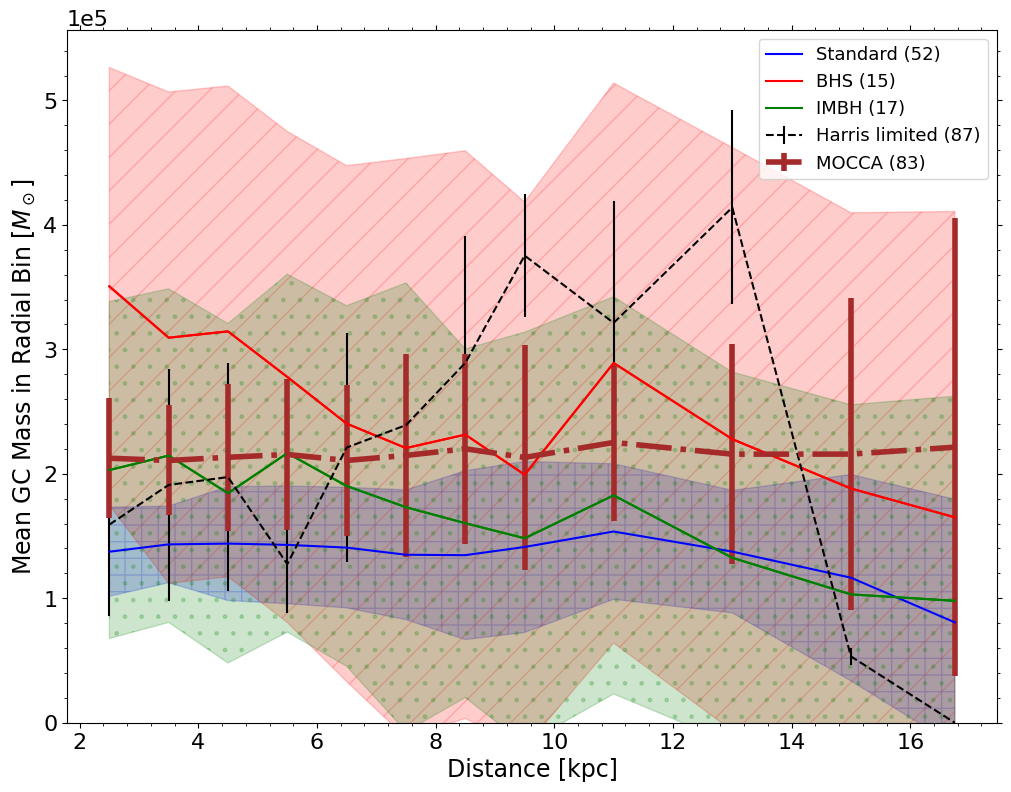}
    \end{subfigure}
    \begin{subfigure}{0.5\textwidth}
       \centering
        \includegraphics[width=\linewidth]{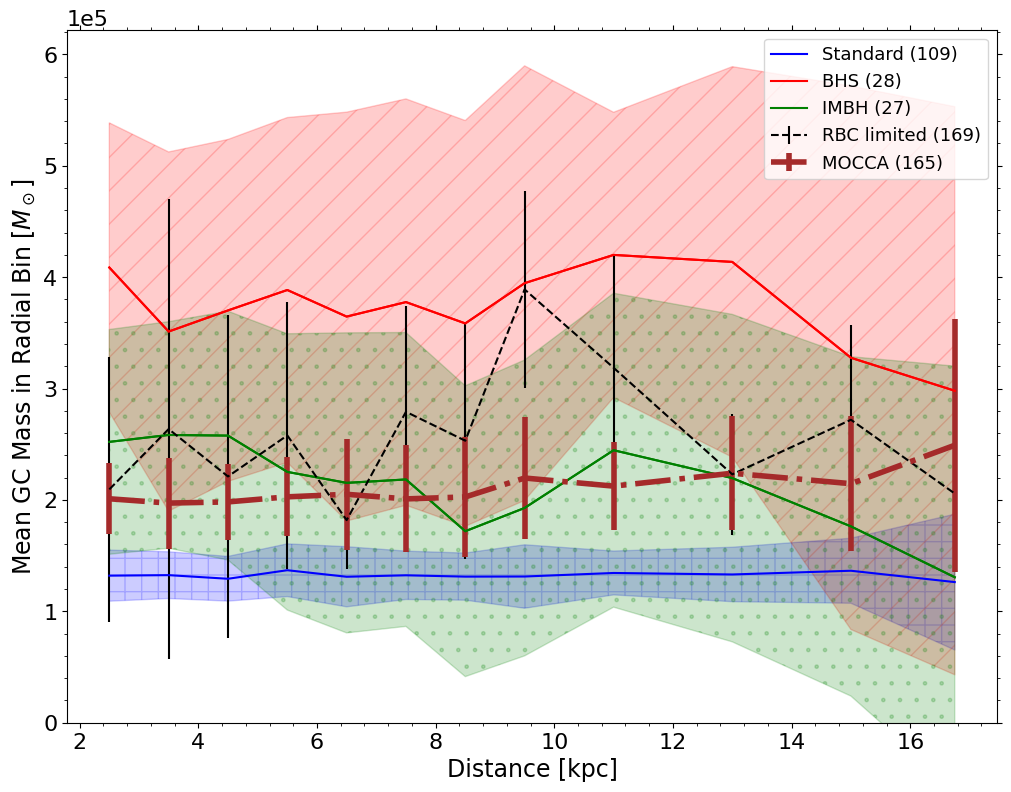}
    \end{subfigure}
    \caption{Mean mass distribution for Standard (blue), BHS (red) and IMBH (green) models for MW (top) and M31 (bottom) for the MOCCA,  respectively. The shaded regions represent the standard deviation for the simulated GC populations. The squared region, the oblique lines and the dots show the standard deviation for Standard, BHS and IMBH models respectively. The mean number of GCs for each dynamical model are reported in brackets. In black dashed lines we report the mean GC mass distributions for the observed populations, while in thick brown dot-dashed line we report the spatial distribution for the entire simulated population.}
    \label{Fig:dynamicalModelMeanMassDistribution}
\end{figure}

Fig. \ref{Fig:dynamicalModelNDistribution} shows the binned radial distribution (total number of models in a given radial bin) of the different dynamical models considered in this study. The Standard model  predominates in the central region of the galactic halo; meanwhile the numbers of dynamical models for all the three dynamical sub-samples seem to be comparable for distance $> 14$ kpc. On the other hand, the BHS models show a higher mean mass distribution at all the galactocentric distances, and IMBH models show a slightly higher mean mass than the Standard models for small galactocentric distances ($< 6-10$ kpc), as it is possible to see in Fig. \ref{Fig:dynamicalModelMeanMassDistribution}. For comparison, the observed number distributions for the MW and M31 are reported in black dashed lines. For the MW population, results from the \citep{Harris1996,Harris2010} catalogue have been used, and for the M31 population the results from the Revised Bologna Catalogue (RBC; \cite{Galletti2004,Galleti2006,Galleti2014}) have been used. For the observational catalogues we applied the same filtering condition carried out in Paper II, that is we limited to GCs within 17 kpc from the galactic center, and with half-light radius surface brightnesses (defined as $L_V / r_h^2$, with $L_V$ being the total $V$ luminosity and $r_h$ the half-light radius) greater than $4000\,\, L_\odot/pc^2$. The distributions reported by our models are comparable with the observed ones.

\begin{figure}
    \centering
    \begin{subfigure}{0.5\textwidth}
       \centering
        \includegraphics[width=\linewidth]{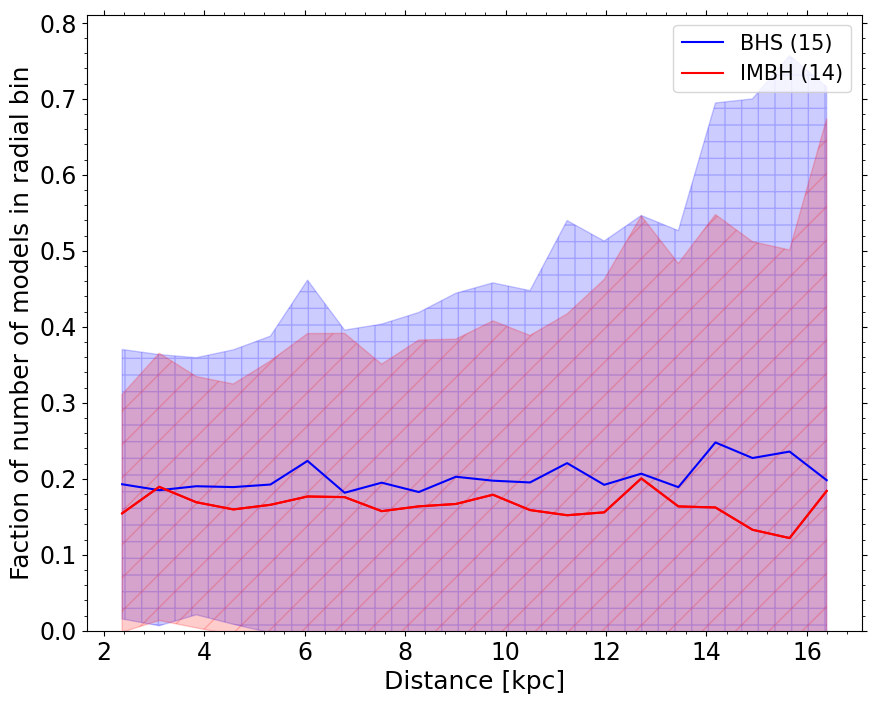}
    \end{subfigure}
    \begin{subfigure}{0.5\textwidth}
       \centering
        \includegraphics[width=\linewidth]{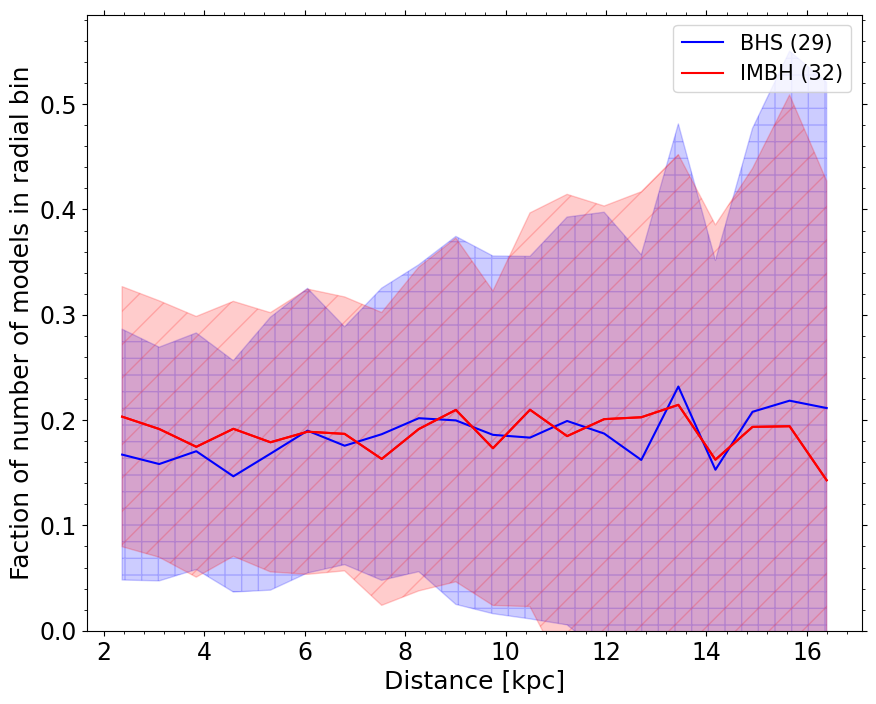}
    \end{subfigure}
    \caption{Spatial distribution for fraction of models in the radial bins for the  BHS (blue) and IMBH (red) models for MW (top) and M31 (bottom) for the MOCCA population, respectively. The shaded regions represent the standard deviation for the simulated GC populations. The oblique lines and the dots show the standard deviation for BHS and IMBH models respectively. The mean number of GCs for each dynamical model are reported in brackets.}
    \label{Fig:dynamicalModelFractionModel}
\end{figure}

In Fig. \ref{Fig:dynamicalModelFractionModel}, we show the radial distribution of the number fraction of BHS and IMBH models in each radial bin. The fraction number has been determined as the mean number of the dynamical models in the radial shell divided by the total number of models in the same shell. It can be seen that the number fraction of both IMBH and BHS models are similar and uniform.

\begin{figure}
    \centering
    \begin{subfigure}{0.5\textwidth}
       \centering
        \includegraphics[width=\linewidth]{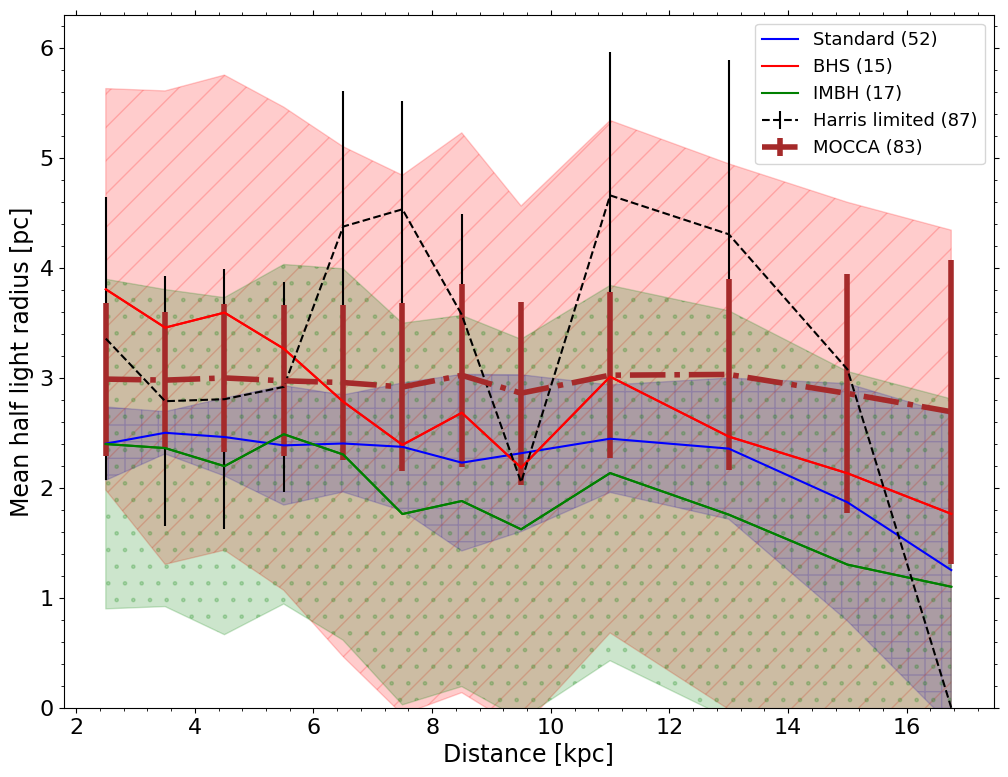}
    \end{subfigure}
    \begin{subfigure}{0.5\textwidth}
       \centering
        \includegraphics[width=\linewidth]{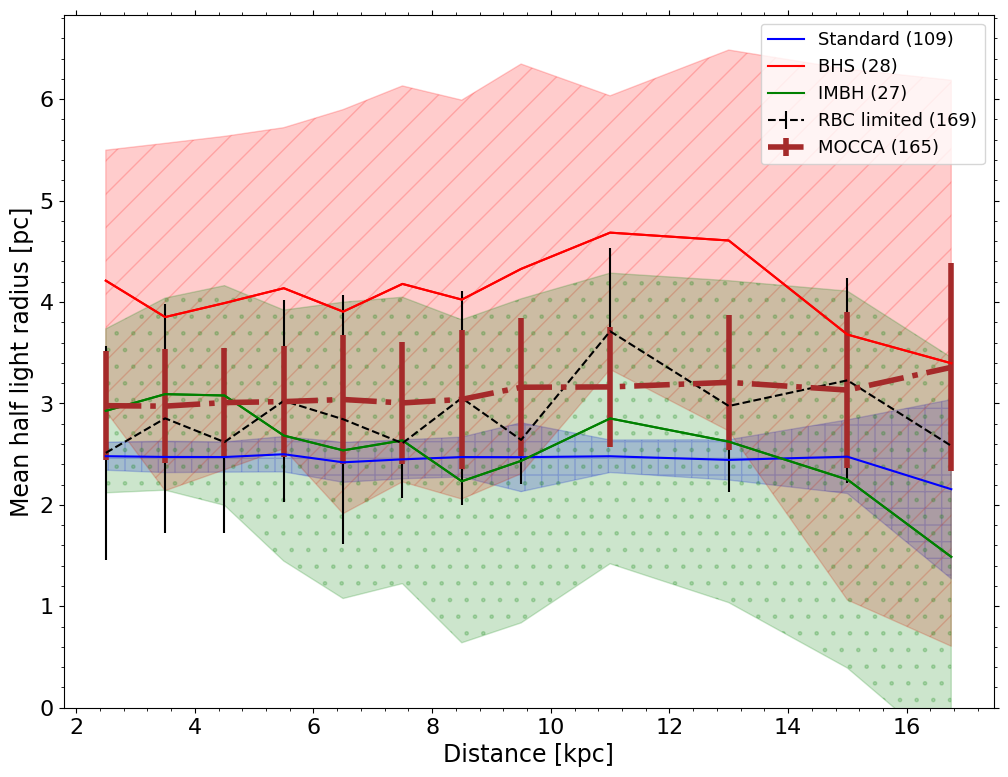}
    \end{subfigure}
    \caption{Mean half-light radius distribution for Standard (blue), BHS (red) and IMBH (green) models for MW (top) and M31 (bottom) for the MOCCA, respectively.The shaded regions represent the standard deviation for the simulated GC populations. The squared region, the oblique lines and the dots show the standard deviation for Standard, BHS and IMBH models respectively. The mean number of GCs for each dynamical model are reported in brackets. In black dashed lines we report the mean half-light radius distributions for the observed populations, while in thick brown dot-dashed line we report the spatial distribution for the entire simulated population.}
    \label{Fig:dynamicalModelMeanRhDistribution}
\end{figure}

In Fig. \ref{Fig:dynamicalModelMeanRhDistribution} we show the mean half-light radius distribution for the different dynamical models for both MW and M31. It is possible to see that for different galactocentric distances in both the MW and M31 populations the mean half light radius of the BHS models is larger compared to Standard and IMBH models, with IMBH models being more compact at larger galactocentric distances. For comparison, the observed number distributions for the MW and M31 are reported in black dashed lines. The observed distributions follow with relatively good agreement the distributions from our simulated models.

\begin{figure}
    \centering
    \begin{subfigure}{0.5\textwidth}
       \centering
        \includegraphics[width=\linewidth]{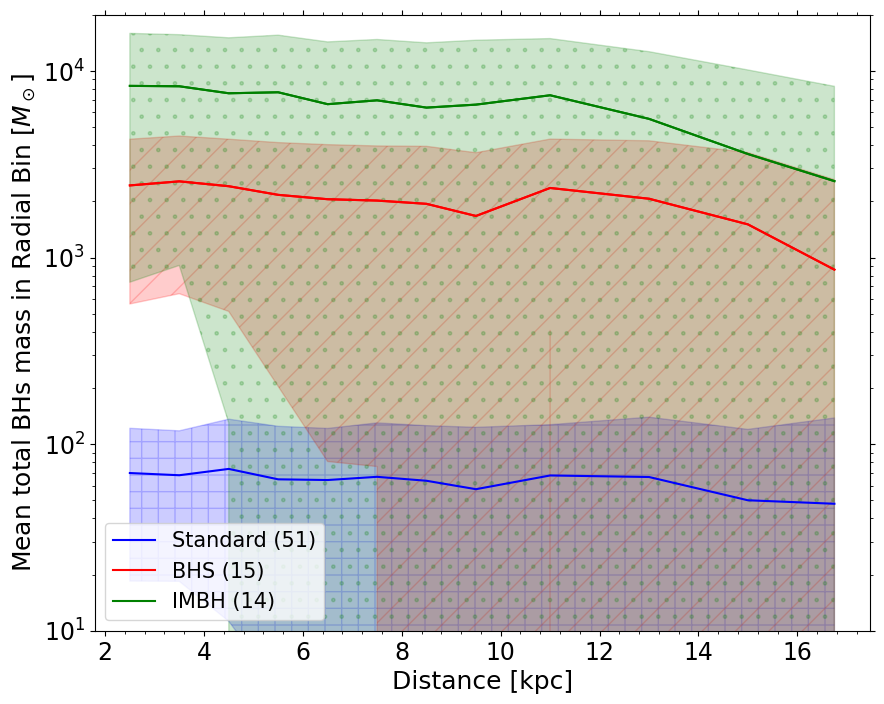}
    \end{subfigure}
    \begin{subfigure}{0.5\textwidth}
       \centering
        \includegraphics[width=\linewidth]{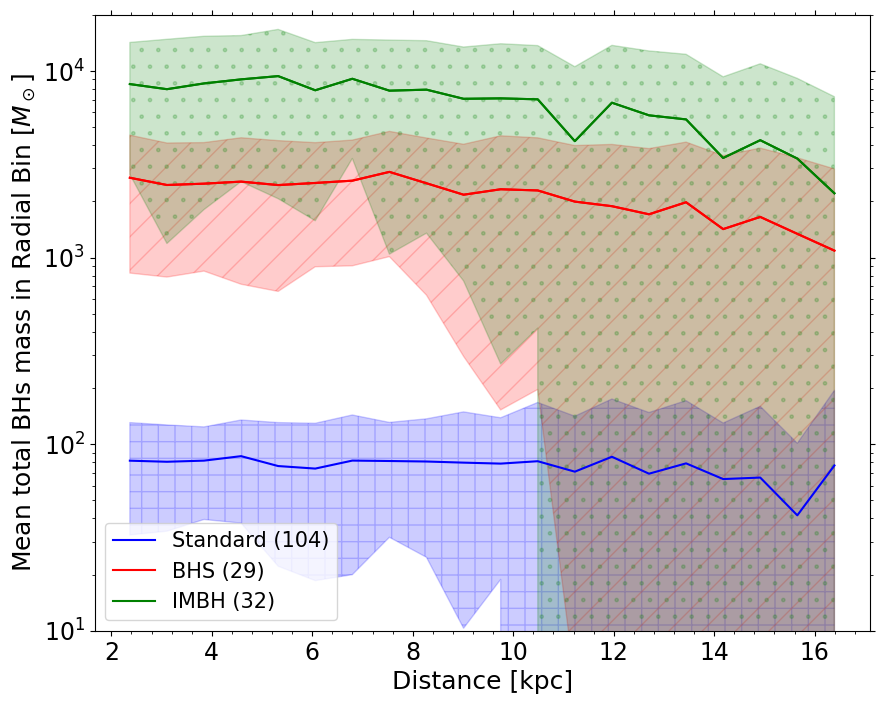}
    \end{subfigure}
    \caption{Mean total BH mass distribution for Standard (blue), BHS (red) and IMBH (green) models for MW (top) and M31 (bottom) for the MOCCA,  respectively. The shaded regions represent the standard deviation for the simulated GC populations. The squared region, the oblique lines and the dots show the standard deviation for Standard, BHS and IMBH models respectively. The mean number of GCs for each dynamical model are reported in brackets.}
    \label{Fig:dynamicalModelMeanMassBHDistribution}
\end{figure}

In Fig. \ref{Fig:dynamicalModelMeanMassBHDistribution} we show the mean total BH mass per GC distribution for the different dynamical models for both MW and M31. The total BH mass per GC is determined as the sum of all BH present in the GC.  The mean total BH mass per GC in the IMBH models is visibly larger in the center of the galactic halo compared to the outskirt ($\sim 4-5$ times larger). This behaviour is a consequence of the fact that in MOCCA models, the most massive IMBHs form in GCs that were born close to galactic center.  These GCs are initially more compact and dense \citep{Giersz2015,ArcaSedda2019} and thus more conducive to forming IMBHs with respect to those in outskirt. However, this is not visible in the BHS model, where the differences in the total BH masses are negligible. Finally, the mean total BH mass in the Standard models is insignificant.

To summarize our results, Standard models are more numerous in the central region of the galactic halo, but they consist typically of low mass and relatively compact GCs, with almost no BHs in the system. Meanwhile the IMBH models show similar global system structure to the Standard models (mean mass and half-light radius), the total BH mass for these models is dominated by the central IMBH. Instead, the BHS models are more massive than both the IMBH and Standard models, and do show a larger half-light radius for almost all galactocentric distances. As already discussed in Paper I, these distributions are correlated with the intrinsic properties of the GCs and their dynamical history, as it will be discussed more in Section \ref{Sec:Discussion}.  The statistical test results are reported in Appendix  \ref{Appendix}.

\subsection{Single and binary BH populations}
The MOCCA-Survey Database I models followed the evolution of the internal dynamics of GCs and also their stellar content, including the BH counts and the BBHs properties.

\subsubsection{Merging binary BH population}
The number of BBHs that would merge in a time range within 12 and 12.5 Gyr and within 10 and 13 Gyr found in our results is presented in Table \ref{Table:BHBHMergers}, together with the number of BBHs that would survive at 12 Gyr. These results consider all the BBH mergers that are generated in GCs, independently of whether the merger actually occurred within the GCs or the binary escaped the GC before merging. For this purpose, also BBHs present in GCs that have merged with the NSC due to dynamical friction are also taken into account. Instead, BBH mergers in dissolved clusters have been excluded from the analysis, together with the mergers between a stellar-mass BH and an IMBH. Considering only a time span within 12 and 12.5 Gyr, our results show that the estimated BBH merger rate expected in 10 years is of the order of $\sim 7 \times10^{-7}$ and  $\sim 10^{-6}$ for MW and M31 respectively. Similar numbers of mergers can be obtained considering a larger time span between 10 and 13 Gyr. The latter time span is more in line with the GCs' age range in the MW and M31, that would vary between 10 and 13 Gyr. Nonetheless, the number of BBH mergers would not differ much between the two time spans. These numbers are reported in Table \ref{Table:BHBHMergers}, and they define the expected BBH merger rate if these populations would be continuously observed for 10 years. The BBHs in GCs can be either primordial, that is the binary formed from the evolution of the two massive stars that were in a binary system in the initial GC model., or dynamically formed, that is the binary formed in the GC via dynamical processes during the GC evolution. The total number of primordial binaries that merged in the two time range considered are reported in Table  \ref{Table:BHBHMergers} too.

\begin{table*}
    \centering
    \begin{tabular}{ c@{\hskip 0.5in} ccccccc}
    \hline
        Galaxy & Present at 12 Gyr & \multicolumn{2}{c}{Merged} &  \multicolumn{2}{c}{Merger rate in 10 yr} & \multicolumn{2}{c}{Primordial Binaries Merged}\\ 
        \hline
         & & 10-13 Gyr & 12-12.5 Gyr & 10-13 Gyr & 12-12.5 Gyr& 10-13 Gyr & 12-12.5 Gyr \\
        \hline
        MW & $120 \pm 16$ (6) & $285 \pm 34$ ($284$) & $30 \pm 6$ ($6$)  & $9.5 \times10^{-7}$ & $6.0 \times10^{-7}$  & $73 \pm 13$  & $4 \pm 2$\\
        M31 & $235\pm 27$ (13) & $515 \pm 50$ ($513$) & $53 \pm 7$ ($7$)& $1.7 \times10^{-6}$&  $1.1 \times10^{-6}$ & $145 \pm 21$ & $5 \pm 2$ \\
    \hline
    \end{tabular}
    \caption{Number of BBHs present at 12 Gyr and merged for two different time ranges and their corresponding merger rate in 10 yr, for MW and M31 respectively. For the BBHs present at 12 Gyr, the total number of primordial binaries are reported in brackets, meanwhile for the merging BBHs, the total number of escaped mergers are reported in the brackets. The number of primordial binaries that merged in the two different time ranges are reported too.}
    \label{Table:BHBHMergers}
\end{table*}

In order to determine the merger rate for BBHs within 1 Gpc, we considered a cosmological cube of a side with a  length of 1 Gpc. Supposing a constant density number of galaxies in the local Universe $\rho_{galaxy}$, the BBH merger rate R in the studied cosmological cube can be determined as 
\begin{equation}
    R = (\rho_{galaxy} \cdot  V \cdot N_{mergers}) / \Delta T,
\end{equation}
with $V$ the volume of the cosmological cube, and $N_{mergers}$ the number of mergers within the time interval $\Delta T = 3$ Gyr. For our study, two different galaxy densities have been used. The  evolution of a cosmological cube with volume $V = 1.2\times10^6\,\rm{Mpc^3}$ and $1.8\times 10^{10}$ particles representing baryonic and dark matter was modelled in Illustris-1 simulation \citep{Vogelsberger2014}. To account for all the bounded galaxy systems present in the simulation at redshift $z=0$, the total number of objects and with mass greater than $10^6\,\,M_\odot$ has been considered, implying a total galaxy density of $\rho_{Illustris} = 0.2\,\rm{ Mpc^{-3}}$. In the local Universe, only 2/3 of galaxies are spirals \citep{Conselice2016}, the total galaxy density for spiral galaxies in the Illustris simulation would be $\rho_{Illustris} = 0.13\,\rm{ Mpc^{-3}}$. Instead, \cite{Abadie2010} estimate the number of accessible Milky Way Equivalent Galaxies (MWEGs) and the extrapolated density of MWEGs in the space, being $\rho_{MWEG} = 0.0116\, \rm{Mpc^{-3}}$. 
We find a merger rate of $R_{Illustris} = 12.7\,\, (22.9)\,\,\rm{yr^{-1}\,\,Gpc^{-3}}$ and $R_{MWEG} = 1.0\,\, (1.8)\,\,\rm{yr^{-1}\,\,Gpc^{-3}}$ in the two cases and for the MW (M31). These results are summarized in Table \ref{Table:BHBHMergeRate}, and will be discussed with  more details in Sec. \ref{Sec:Discussion}.

\begin{table}
    \centering
    \begin{tabular}{ ccc}
    \hline
        Galaxy & $R_{Illustris}\,\,(D=1\,\rm{Gpc})$ [$\rm{yr^{-1}}$] & $R_{MWEG}\,\,(D=1\,\rm{Gpc})$ [$\rm{yr^{-1}}$]\\
        \hline
        MW  & 12.7 &  1.0\\
        M31 & 22.9 &  1.8\\ 
    \hline
    \end{tabular}
    \caption{The merger rate $R$ for MW-like galaxies within a distance of 1 Gpc, using the galaxy density from Illustris and the interpolated density of MWEGs, for both MW and M31 respectively.}
    \label{Table:BHBHMergeRate}
\end{table}

\begin{figure*}
    \centering
    \includegraphics[width=\linewidth]{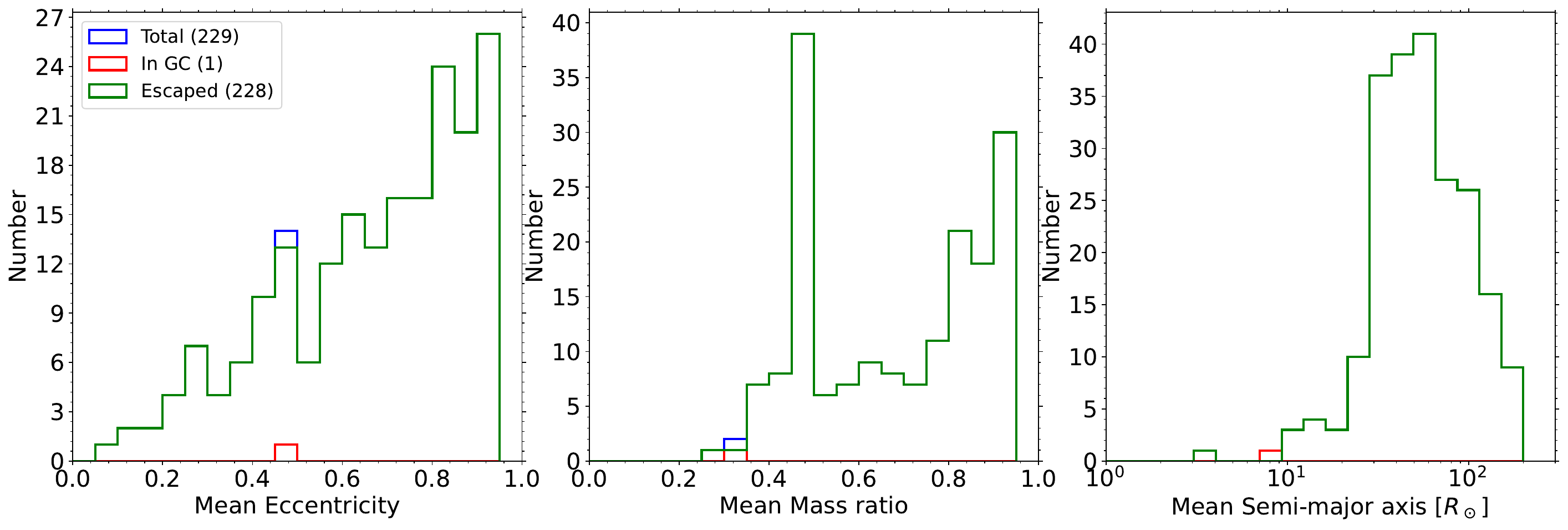}
    \caption{Semi-major axis  (left), orbital eccentricity (middle) and mass ratio (left) histograms for the BBHs that would merge in the time range between 10 and 13 Gyr in the MW population. The distribution for the BBHs that would merge in the GC or that escaped the GC when merged are reported. In brackets the number of each sample has been reported.}
    \label{Fig:BHBHMergedHistrograms}
\end{figure*}

\begin{figure*}
    \centering
    \includegraphics[width=\linewidth]{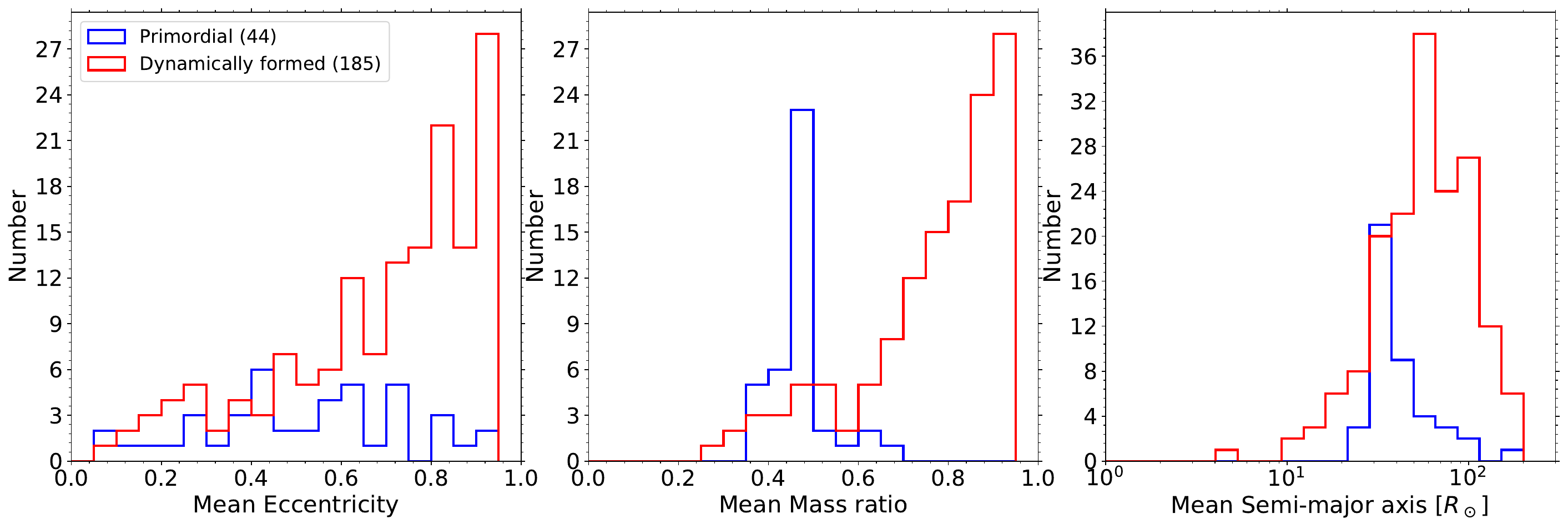}
    \caption{Semi-major axis  (left), orbital eccentricity (middle) and mass ratio (left) histograms for the BBHs that would merge in the time range between 10 and 13 Gyr in the MW population. The distribution for primordial and dynamically formed BBHs that would merge are reported. In brackets the number of each sample has been reported.}
    \label{Fig:BHBHPrimordial}
\end{figure*}

The semi-major axis, eccentricity and mass ratio of the BBHs that would merge in the time range within 10 and 13 Gyr are reported in Fig. \ref{Fig:BHBHMergedHistrograms}, together with the distribution of BBHs that merged in the GC or that escaped the host GC at the merging time. As it is possible to note, binaries with high eccentricity and small semi-major axes (< 100 $R_\odot$) would merge in this time range. Also, most of the merged binaries have a high mass ratio, meaning that the mass difference of the two BHs are negligible.  The mass ratio reproduced in our simulations differs from the value observed in LIGO/VIRGO BBH mergers. As mentioned already before, these differences are expected due to the outdated BH mass prescription used in the MOCCA-Survey Database I. Instead, the distribution properties of the merger BBHs for primordial and dynamically formed binaries are shown in Fig. \ref{Fig:BHBHPrimordial}. The dynamically formed binaries have more eccentric orbit compared to the primordial ones, and also have a larger mass ratio. Also, the semi-major axis for the dynamically formed binaries seems to be larger than the primordial binaries.

\subsubsection{Non-merging binary BHs}
\begin{figure*}
    \centering
    \begin{subfigure}{\columnwidth}
       \centering
        \includegraphics[width=\linewidth]{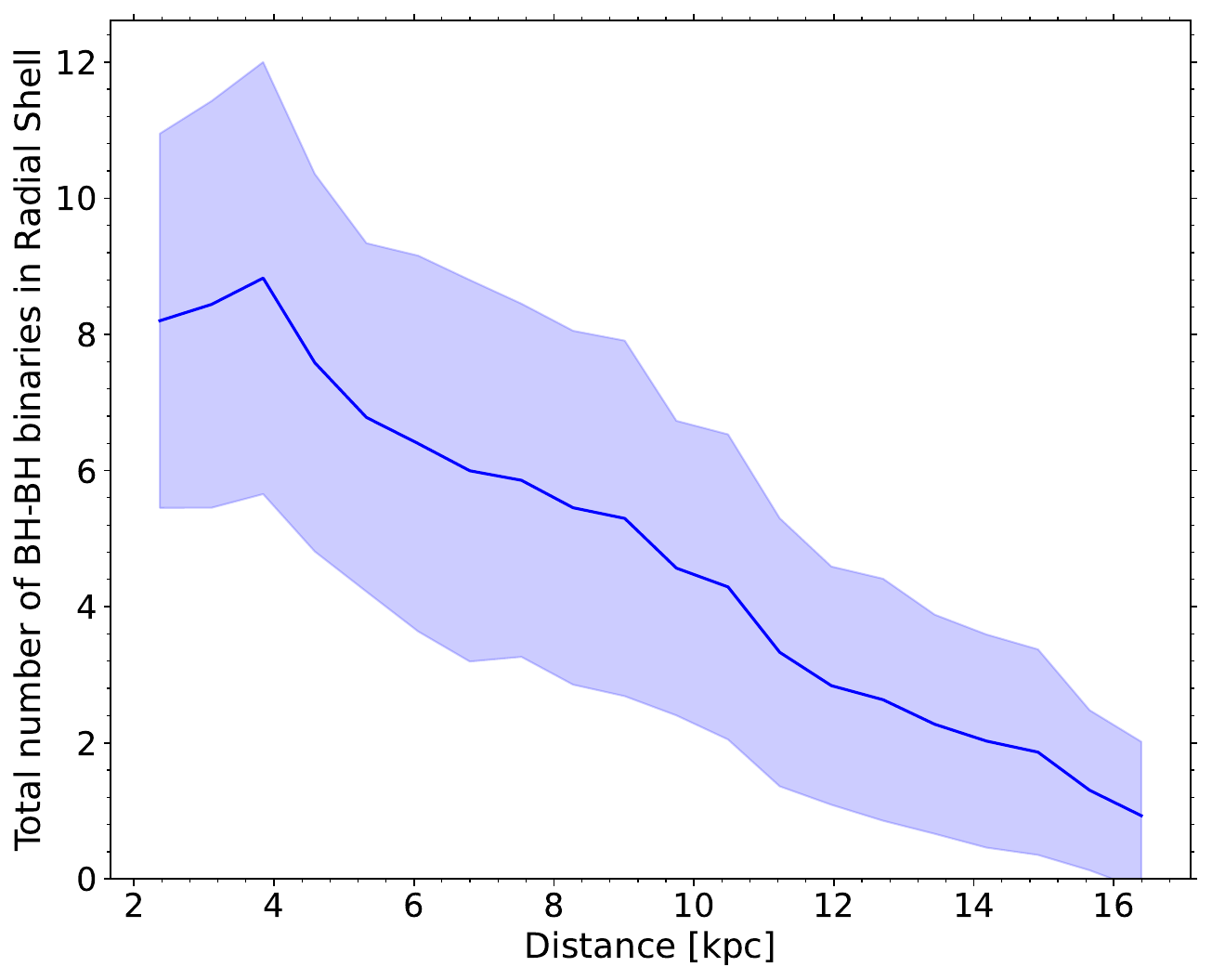}\\
        \hfill
        \vspace{10.5pt}
        \centering
        \includegraphics[width=\linewidth]{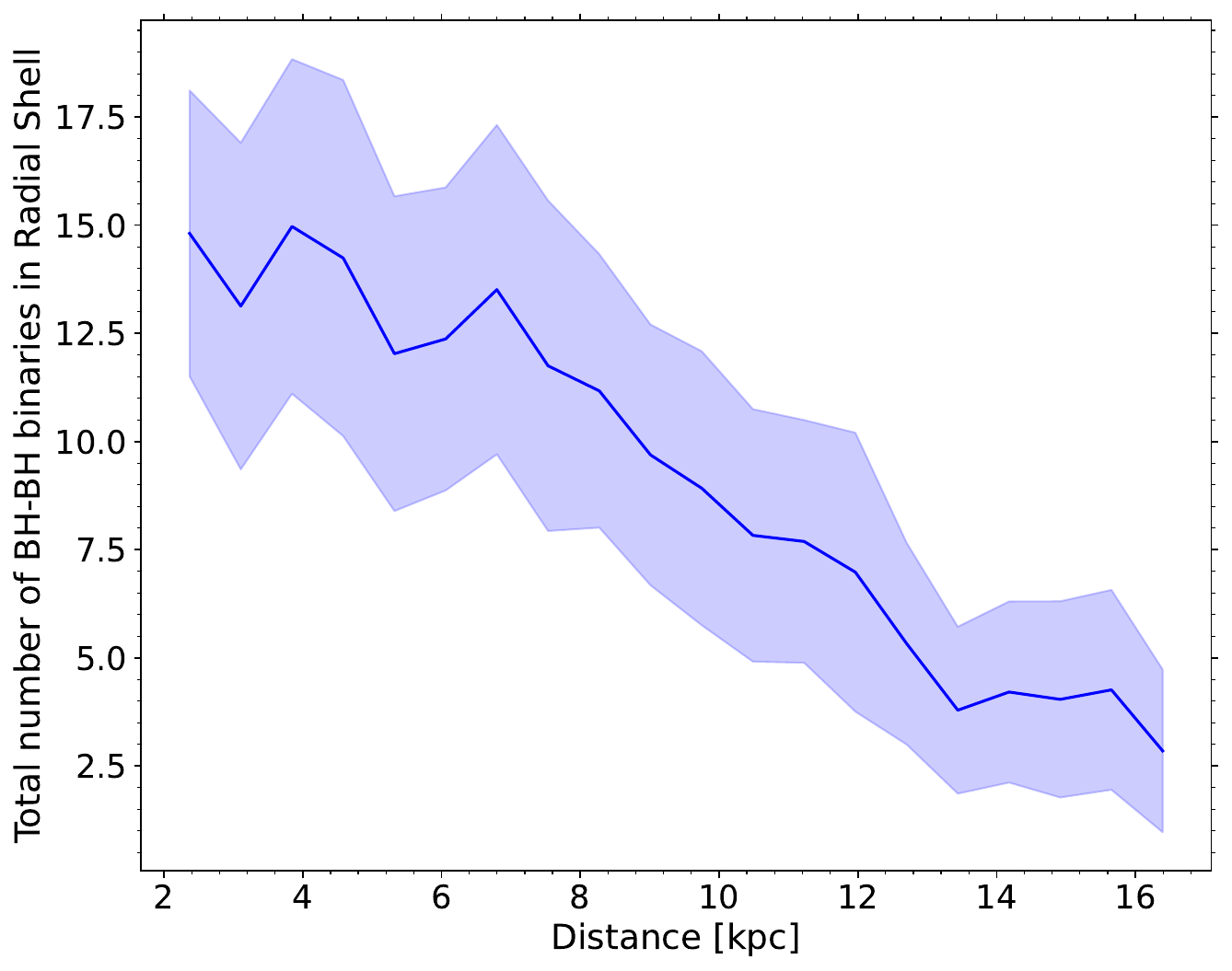}
    \end{subfigure}
    \begin{subfigure}{\columnwidth}
       \centering
        \includegraphics[width=\linewidth]{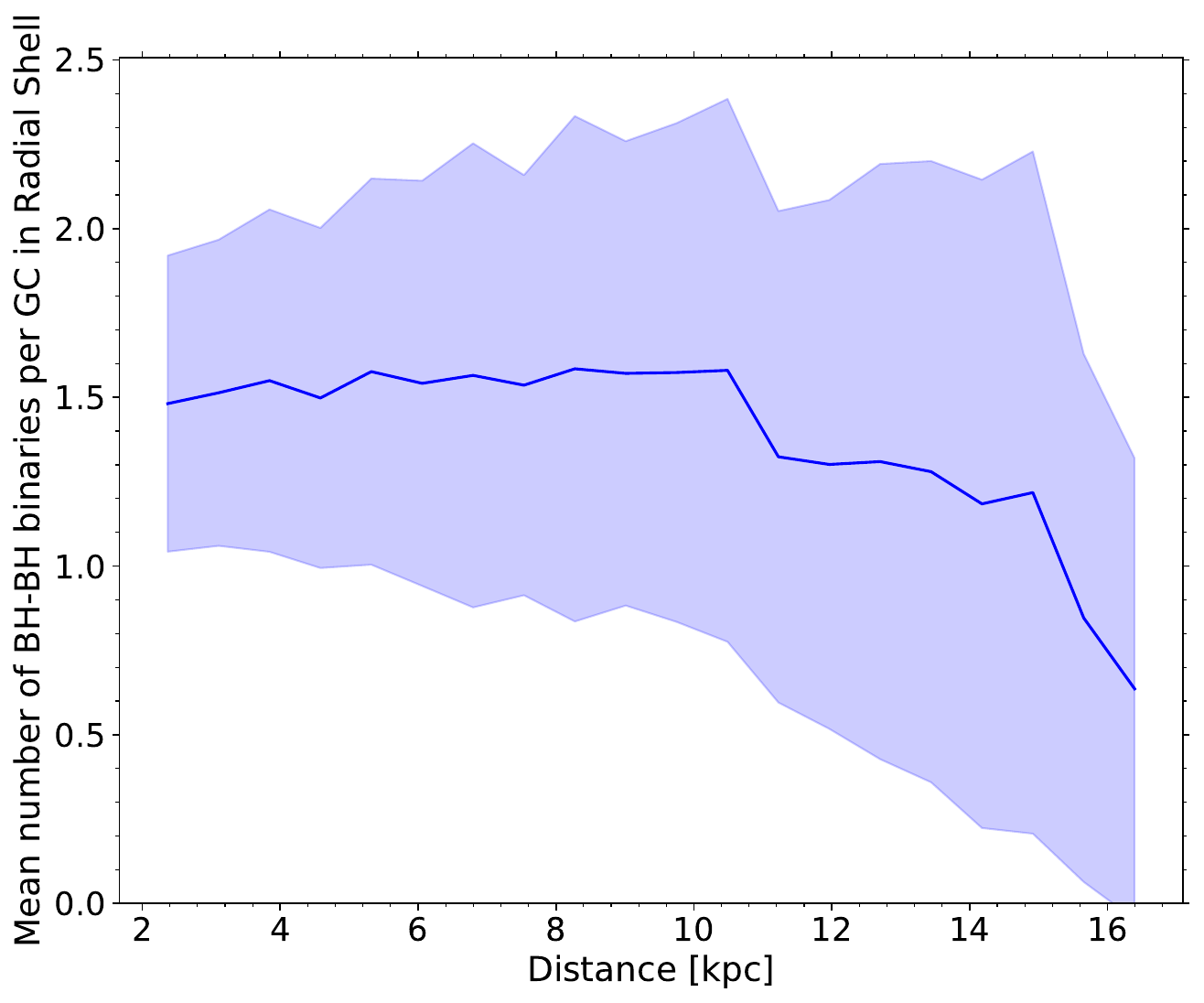}\\
        \hfill
        \vspace{-1pt}
        \centering
        \includegraphics[width=\linewidth]{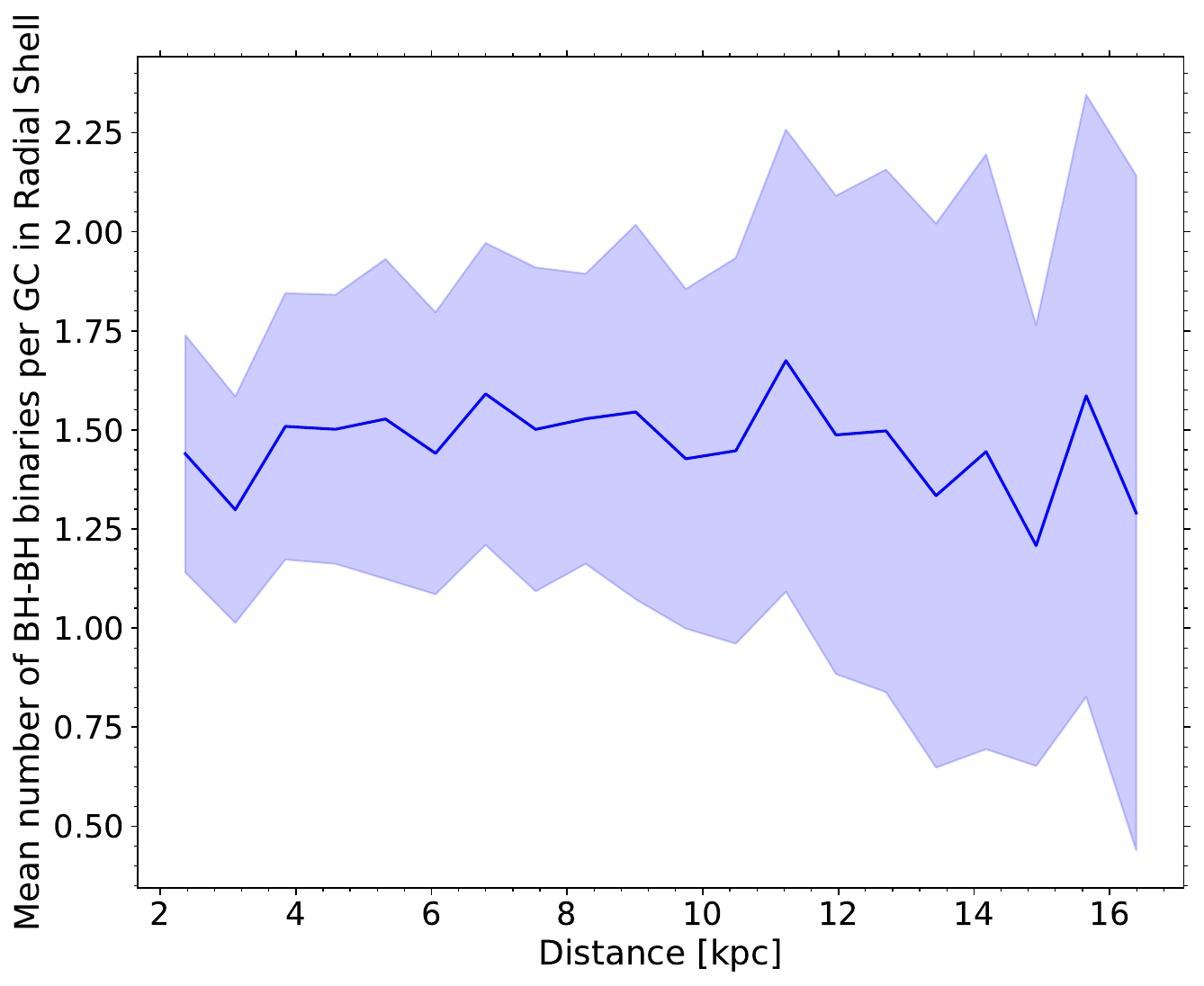}
    \end{subfigure}
    \caption{Total number (left) and mean number (right) of non-merging BBH distribution for MW (top) and M31 (bottom) for the MOCCA,  respectively. The shaded regions represent the standard deviation for both the observed and the simulated GC populations.}
    \label{Fig:BHBHnumberDenistyDistribution}
\end{figure*}

The  distributions for the total number and the mean number of BBHs at 12 Gyr at different galactocentric distances are shown in Fig. \ref{Fig:BHBHnumberDenistyDistribution} for both MW and M31 respectively. The mean number of BBHs has been determined as the total number of BBHs in the radial bin divided by the total number of GCs in the same radial shell. Our results show that most of the BBHs are found in the central region of the galaxy. However, the mean number of BBHs per GC is constant for different galactocentric distances. Indeed, the p-values for the Kolgomorov-Smirnov test (KS test) comparing our results with a uniform distribution in galactocentric distances are 0.68 and 0.64 for MW and M31 respectively, both with the ``two-sample'' alternative hypothesis. This implies that the mean number of BBHs per GC are uniformly distributed in galactocentric radius.

\begin{figure*}
    \centering
    \begin{subfigure}{\columnwidth}
       \centering
        \includegraphics[width=\columnwidth]{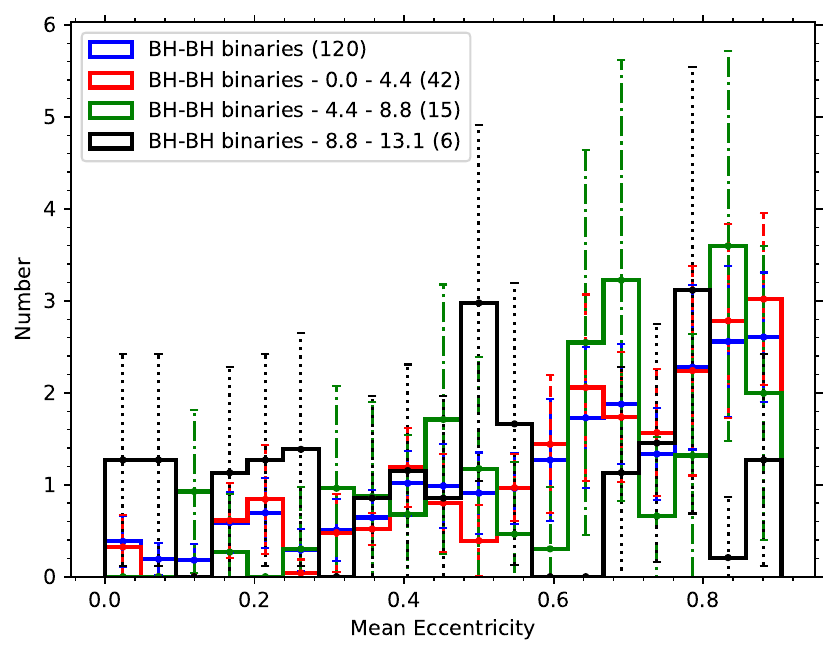}
    \end{subfigure}
    \begin{subfigure}{\columnwidth}
       \centering
        \includegraphics[width=1.05\columnwidth]{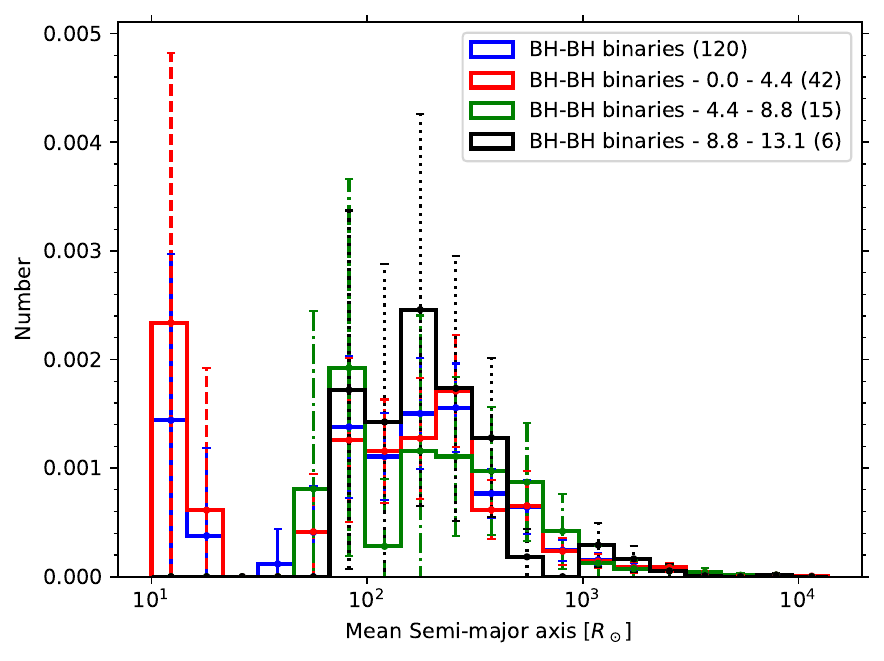}
    \end{subfigure}
    \includegraphics[width=\columnwidth]{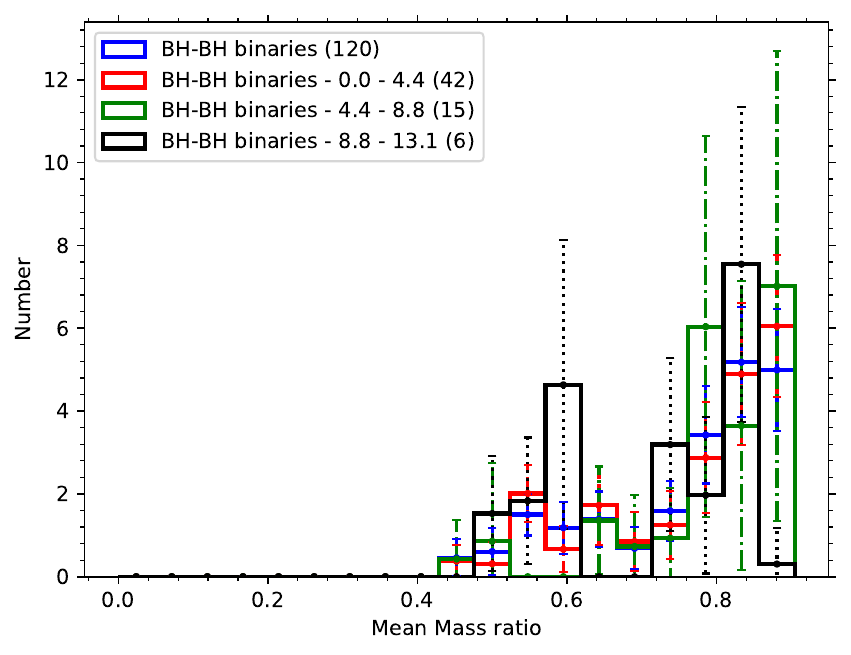}
    \caption{Orbital eccentricity (top-left), semi-major axis (top-right) and mass ratio (bottom) histograms for the BBHs that are present at 12 Gyr in the MW population. The distribution for all BBHs (blue), and for different galactocentric distance shells are reported. The mean number of BBHs for each population is reported in brackets. The area beneath the histograms have been set to  1.}
    \label{Fig:BHBHSurvivedPropMW}
\end{figure*}

\begin{figure*}
    \centering
    \begin{subfigure}{\columnwidth}
       \centering
        \includegraphics[width=\columnwidth]{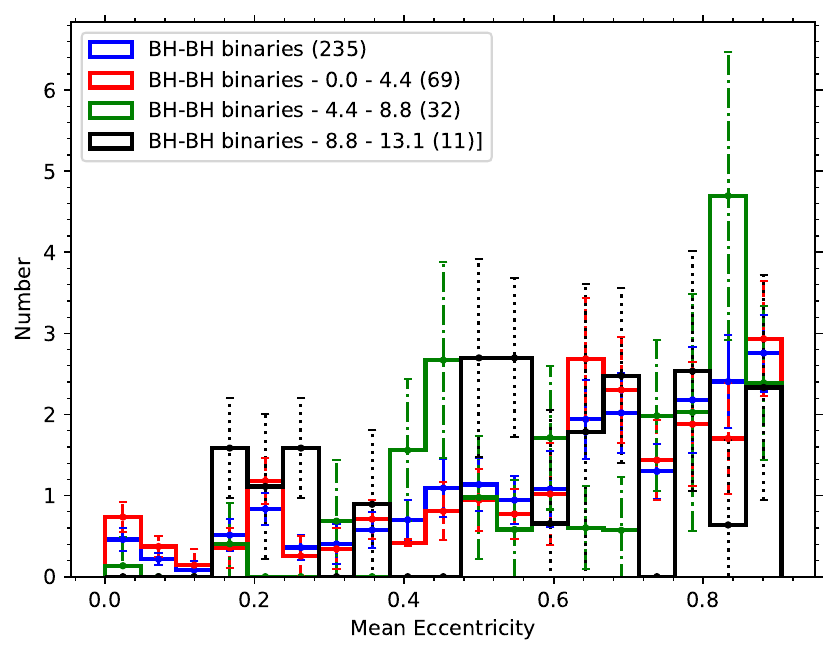}
    \end{subfigure}
    \begin{subfigure}{\columnwidth}
       \centering
        \includegraphics[width=1.1\columnwidth]{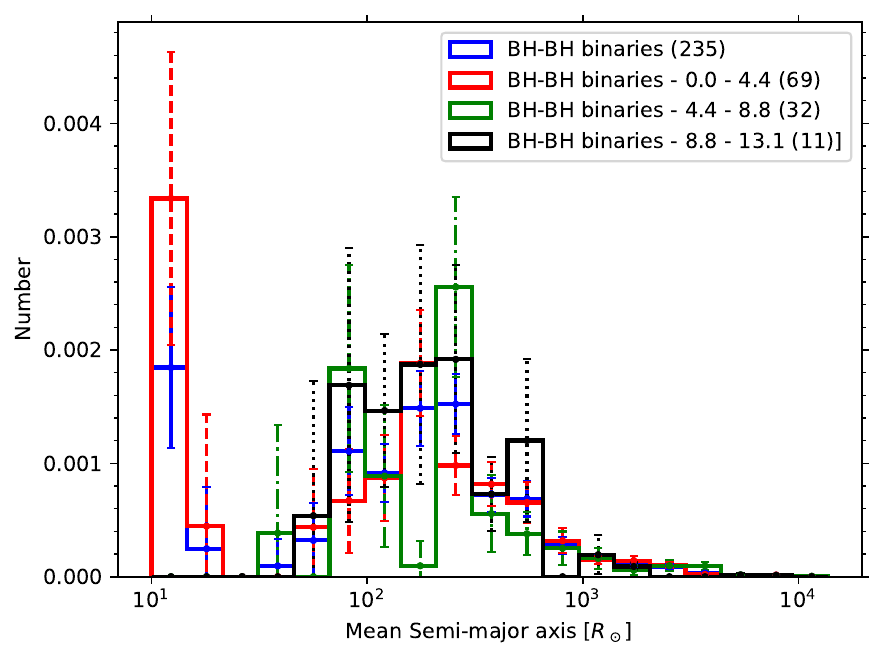}
    \end{subfigure}
    \includegraphics[width=\columnwidth]{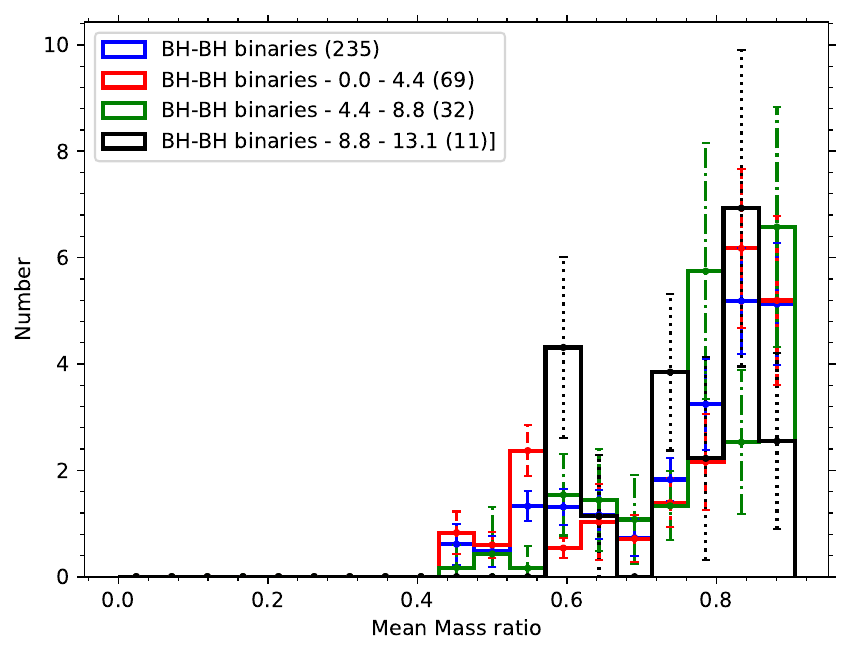}
    \caption{Orbital eccentricity (top-left), semi-major axis (top-right) and mass ratio (bottom) histograms for the BBHs that are present at 12 Gyr in the M31 population. The distribution for all BBHs (blue), and for different galactocentric distance shells are reported. The mean number of BBHs for each population is reported in brackets. The area beneath the histograms have been set to  1.}
    \label{Fig:BHBHSurvivedPropM31}
\end{figure*}

The mean number of non-merging BBHs found at 12 Gyr in our simulations is $120\pm 16$ and $235\pm 27$ for MW and M31 respectively. The normalized histograms (the area beneath the histograms have been set to 1) of the orbital eccentricity, mass ratio and semi-major axis of the BBHs at 12 Gyr are reported in Figs.  \ref{Fig:BHBHSurvivedPropMW} and \ref{Fig:BHBHSurvivedPropM31} for MW and M31 respectively. The orbital eccentricity of the BBHs are mostly concentrated in two regions: extremely eccentric ($\sim 1.0$) and almost circular ($\sim 0.2$), whereas the semi-major axis of the binaries are relatively compact with a mean value of $\sim 50 \,\,R_\odot$). On the other hand, the mass ratio of the binaries is mostly concentrated in the region between 0.8 and 1. In the histograms, the distribution for different galactocentric distance shells are reported too. As it is possible to see, the orbital eccentricity of BBHs is more extended to circular orbit at larger galactocentric distances, with large part of the population having a more thermal orbital eccentricity ($> 0.6$) in the central galactic regions. Similarly, the mass ratio and the orbital semi-major axis at larger galactocentric distances seem to be more extended towards small values ($<0.6$)  and larger values ($>10^2 R_\odot$) respectively. Instead, for smaller galactocentric distances, the mass ratio peaks for larger values ($>0.8$), with the semi-major axis peaking at smaller values ($<10^2 R_\odot$).

\begin{figure*}
    \centering
    \includegraphics[width=\linewidth]{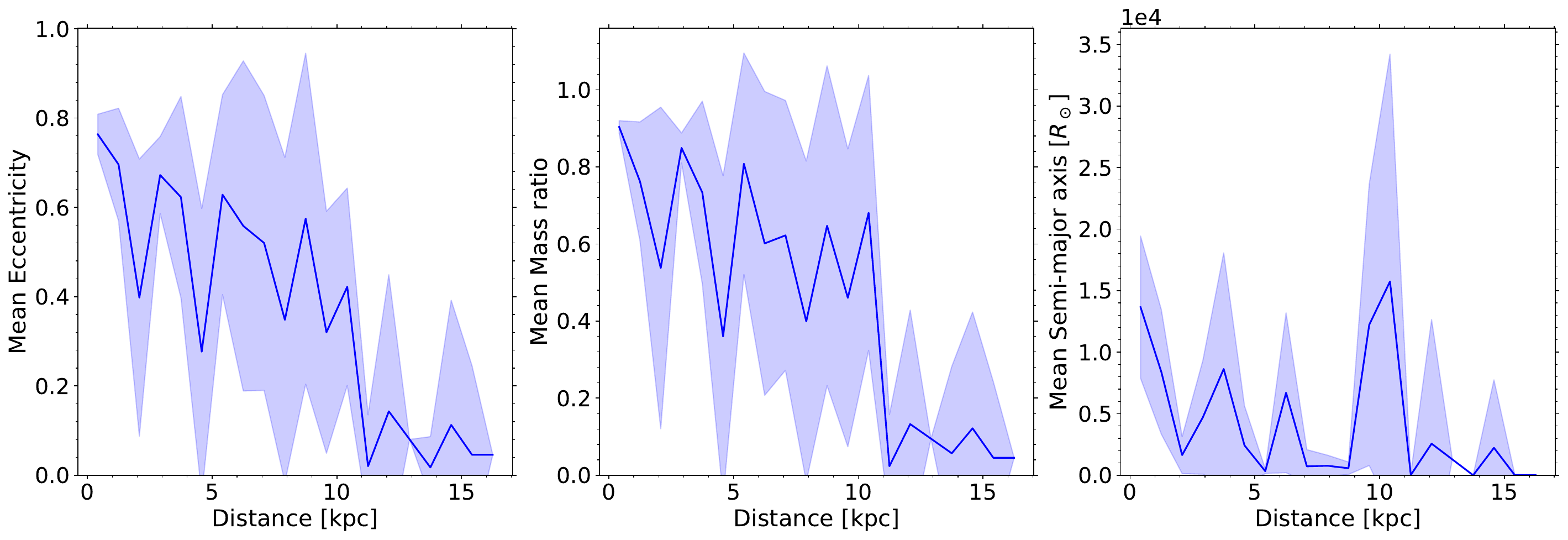}
    \caption{Distributions for mean values in radial galactocentric bins for semi-major axis  (left), orbital eccentricity (middle) and mass ratio (left) histograms for the BBHs that are present at 12 in the MW population. The shaded regions represent the standard deviation for both the observed and the simulated GC populations.}
    \label{Fig:BHBHSurvivedRadialPropMW}
\end{figure*}

The spatial distribution for the semi-major axis, eccentricity and mass ratios are showed in Fig. \ref{Fig:BHBHSurvivedRadialPropMW} for the MW BBHs present at 12 Gyr. We found that the distributions are a statistically decreasing function with galactocentric distance (the statistical test results are reported in Appendix  \ref{Appendix}).  These results show that close BBHs might be found in the central region of the galactic halo, where they are also more numerous. Also, this implies that it might be expected that the number of BBH mergers are more likely to be observed at smaller galactocentric distances.

\subsubsection{BH delivered to the NSC}
\begin{figure}
    \centering
    \begin{subfigure}{0.5\textwidth}
       \centering
        \includegraphics[width=\linewidth]{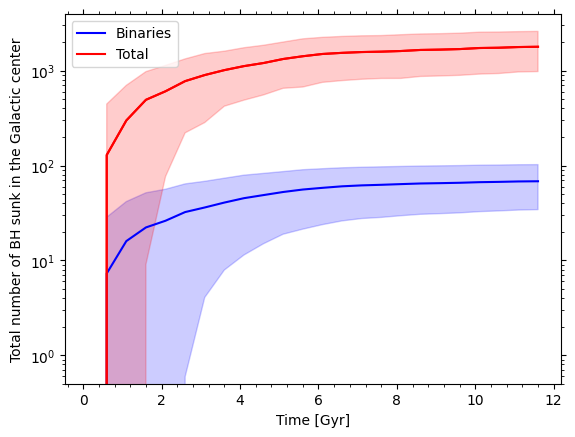}
    \end{subfigure}
    \begin{subfigure}{0.5\textwidth}
       \centering
        \includegraphics[width=\linewidth]{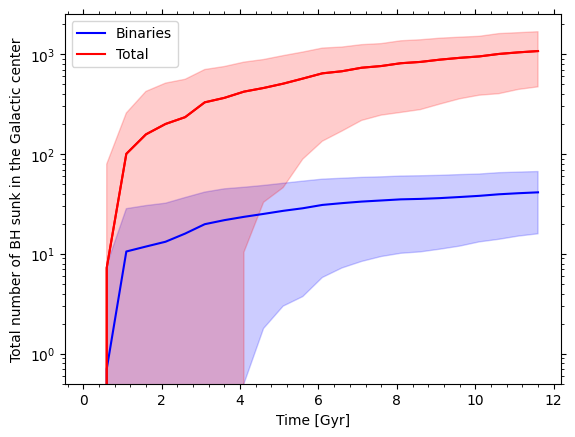}
    \end{subfigure}
    \caption{Total number of BH and BBHs delivered to the NSC time evolution for MW (top) and M31 (bottom) respectively.}
    \label{Fig:BHDeliveredNSC}
\end{figure}

The evolution of the total number of BHs delivered to the NSC are reported in Fig. \ref{Fig:BHDeliveredNSC}. It can be seen that a significant number of BHs and BBHs have been delivered to the NSC by GCs within a few Gyr after their formation ($\sim 30\%$ of the total number of BHs are delivered in the first 2 Gyr). A slow increase in these numbers is seen at later times. The total number of binaries delivered to the NSC is $\sim 5\%$ of the total BH population that were delivered to the NSC. The mean total number of BH delivered to the NSC are $\sim 3000$ and $\sim 1000$ for MW and M31 respectively, of which $\sim 100$ and $\sim 60$ are BBHs for MW and M31 respectively.

For an initial power-law GCIMF between $M_{low} = 10^3-10^4\,M_\odot$ and $M_{up}=10^7\,M_\odot$, the initial total number of GCs in the population would be 10 to 5 times larger respectively, and a total mass of the GCs population $\sim 6-5$ times larger. Because of the fast cluster dissolution time for low mass clusters at small galactocentric distances \citep{Gnedin1997,ArcaSedda2014,Rodriguez2022}, we would expect that the number of GCs delivered to the NSC due to dynamical friction would be 3-4 times larger only, with the total mass of roughly one order of magnitude larger. This will imply that the number of BHs and BBHs delivered to the NSC of few times larger only.

\section{Discussion} \label{Sec:Discussion}
The kinematic comparison with the MW GC population in the Bajkova catalogue \citep{Bajkova2021} shows that the GC population simulated in our models represent decently the observed kinematic properties. To model the Galactic potential, the MOCCA models were simulated using a point mass approximation and GCs were assumed to move on a circular orbit. From Fig. \ref{Fig:densityMapBajkowaRcMassDistMW} it is possible to note that the orbital approximations deployed in the MOCCA models is in relatively good agreement with the observational data in the MW GC population, within the simulated mass range. This is a crucial and important result for our models: indeed, the GCs in our machinery were initially placed in a circular orbit around the external galaxy and then they were modelled in elliptical orbits, given the findings in \cite{Cai2016}. Nonetheless, despite the limitations that this assumption would imply, the distribution of the observed GC mass and their circular orbits are in agreement with our simulations.

The distribution of the mean GC mass for different dynamical models is similar between the MW and M31 population. The BHS models tend to be more massive than the Standard and IMBH models, and the IMBH models being more massive than the Standard ones. These results would suggest that the most massive GCs might contain a BHS in their center. Similarly, some correlation between the BHS mass and the galactocentric distance might exist, with more massive BHS GCs seen at small galactocentric distances.

 As already shown in Paper I, the BHS models are expected to have a larger half light radius since the central energy generation (controlled by the BHs) is much stronger compared to the other systems, implying a more expanded system. On the other hand, it is expected that the influence of the IMBH would change the central properties of the GC: due to the deeper central potential, the system is expected to be more concentrated, implying a smaller half light radius. This is indeed seen in our simulations for different galactocentric distances in the both MW and M31 populations: the mean half light radius of the BHS models is larger than the Standard and IMBH models, with IMBH models being more compact at larger galactocentric distances. In comparison with the results shown in Paper I, the results shown in this work also take into consideration the interaction between the GC and the host galaxy in the survival of the GC itself. This might be a further support to our machinery results and assumptions.

The mean total BH mass in the system is significantly larger for the IMBH model compared to the BHS and Standard ones. For these models, the total BH mass in the GC is defined by the mass of the IMBH. Indeed, the presence of the IMBH would imply a high density and short dynamical interaction time-scale that would drive out all the massive BHs from the system. Also, the mean total BH mass in the GC for the IMBH models is larger at a smaller galactocentric distance. This might imply a correlation between the formation of an IMBH in a GC and the galactocentric distance and the local galactic density. Instead, the number of BHs in the Standard models are expected to be small (if any), meaning that it is expected to have an almost null BH mass in the system. 

\cite{Lutzgendorf2013} examined for the existence of a possible IMBH at the center of the 14 GCs in their sample using observed surface brightness profiles and velocity dispersion profiles. Six of them have been proposed by \cite{Lutzgendorf2013} to host an IMBH in their center (NGC 1904, NGC 5139, NGC 5286, NGC 6266, NGC 6388, and NGC 6715), and among these, 4 GCs have a galactocentric distance greater than 5 kpc. The number density of GCs decreases for galacocentric distances greater than about  5 kpc, with a fewer number of GCs observed at larger distances. For this reason, we have set a minimum galactocentric distance of 5 kpc for this comparison. \cite{ArcaSedda2019} classified Galactic GCs IMBH (or BHS) based on how many MOCCA models had a BHS or a IMBH, finding a total of 35 models harbouring an IMBH. 16 of the 35 IMBH reported Galactic GCs in  \cite{ArcaSedda2019} are found a distances $ > 5$ kpc.

 \cite{Askar2018} and \cite{ArcaSedda2019} reported the number of GCs candidates that would harbour a BHS in their center, both using the MOCCA-Survey Database I results to identify BHS models. In \cite{Askar2018}, the authors chose models based on their central surface brightness and the observed current half-mass relaxation time, with a total of 28 Galactic BHS GCs. \cite{ArcaSedda2019} reported a total of 23 models harbouring a BHS. Instead, the authors of \cite{Weatherford2019} found that the mass segregation parameter $\Delta$, which was derived from the 2D-projected snapshots of the models published in the CMC Cluster Catalogue \citep{Kremer2020} correlates with the number of BHs in the system. A total of 29 Galactic BHS GCs were found in their work (considering only GCs that retain a number of BH $N_{BH} > 50$). More than half of the reported GCs in \cite{Askar2018}, \cite{ArcaSedda2018} ($\sim 70\%$) and in \cite{Weatherford2019} are found at distance $> 5$ kpc. From our models, it might be expected that the probability to discover an IMBH or BHS model is higher at larger galactocentric distances. Indeed, the total number of Standard models seems to be comparable with the IMBH and BHS ones in the outskirts of the galactic halo: the percentage over the whole population of GCs with galactocentric distance $>5$ kpc is $\sim 20\%$ for both IMBH and BHS models. The mean IMBH mass decreases with the galactocentric distances, suggesting a link for the IMBH formation to the galactic field and galactocentric position. However, this is not seen for the BHS models.

The simulated merger rate  within a volume $V = 1\,\rm{Gpc^3}$ obtained by our simulations depends on the galaxy density number assumed. Considering the galaxy density number from the Illustris simulations \citep{Vogelsberger2014}, we have a merger rate of $\sim 13-23\,\,\rm{yr^{-1}\,\,Gpc^{-3}}$, meanwhile considering the interpolated density of MWEGs from \cite{Abadie2010}, we obtain a merger rate of $\sim 1.0-2.0\,\,\rm{yr^{-1}\,\,Gpc^{-3}}$. The results from the Illustris simulation could be interpreted as maximum expected merger rate. Indeed, the galaxy density number used in this work  assumes that all galaxies (particularly dwarf ones) in the cosmological cube considered in our study are MW- and M31-like galaxies. On the other hand, the results from the interpolated density of MWEGs should be considered as a minimum expected merger rate, due to the set of assumptions that have been made.  Indeed, the merger rates for both MW and M31 in our simulations have been determined for a sub-sample of the whole GC population. As reported in Paper II, we constrained our study for GCs with initial masses between $2\times10^5$ and $1.1\times 10^6 M_\odot$, and within 17 kpc from the galactic center. The most massive GCs are actually excluded in our study. These  might be expected to be host to a large number of BBHs and in particular to BBHs mergers. Also, one additional explanation for the smaller reported value in our simulation is the prescription used in the MOCCA-Survey Database I. As said before, the BH masses obtained in the old prescription are smaller compared to the new updated ones, leading to a larger GW decay time in our simulations. Finally, in our study we did not consider all models that initially populated the studied galaxies. Indeed, we have considered only GCs that would survive up to 12 Gyr, with the dissolved GCs not taken into account. The latter would contribute importantly to the merger rate.

Nonetheless the assumption made in the determination of our merger rate, our values are comparable to the one reported in \cite{Mapelli2022}, where a local merger rate $R \sim 4-8 \,\,\rm{yr^{-1}\,\,Gpc^{-3}}$ have been reported. In their analysis, \cite{Mapelli2022} explored the cosmic evolution of isolated BBHs and dynamically formed  BBHs in nuclear star clusters, GC and young star clusters. In particular, they studied the BBH merger rate considering two main different supernova channels \citep{Fryer2012} (that are rapid and delayed models) in the BH formation, finding different results according to the chosen prescription: the merger rate in the delayed models are roughly $40-60\%$ smaller compared to the merger rate in the rapid one. The different merger rate can be explained by the different minimum BH mass in the two models ($5 M_\odot$ for rapid and $3 M_\odot$ for the delayed), leading to a larger GW decay time in the delayed model compared to the rapid ones.

\cite{Banerjee2022} investigated the importance of binary evolution and cluster dynamics in producing merging BBHs over cosmic time. The author performed a population synthesis for the modelled universe, deploying direct N-body simulations to model the evolution of young massive clusters, and stellar evolutionary models for isolated binaries.  The author's estimates of the intrinsic BBH merger rate density and the cosmic evolution are consistent with the findings in the GWTC-2. In particular, the intrinsic merger rate for BBHs considering only  young massive clusters determined in \cite{Banerjee2022} is of the order $\sim 1 \,\,\rm{yr^{-1}\,\,Gpc^{-3}}$ for redshift $z=0$. This value is comparable to  our results, with most of the models computed in \cite{Banerjee2022} used the rapid supernova prescription.

Finally, \cite{Askar2017} determined the local merger rate density using models from the MOCCA-Database I. Assuming a  GC star formation rate from \cite{Katz2013}, the author determined local merger rate density as function of the star formation rate and the probability of forming a BBH per unit delay time. The authors find a merger rate $R = 5.4  \,\,\rm{yr^{-1}\,\,Gpc^{-3}}$. Our results are in relatively good agreement with the value reported in \cite{Askar2017}. Also, our estimates place in the ballpark of Ligo-Virgo Collaboration predictions for BBH merger rate of  $R = 23.9  \,\,\rm{yr^{-1}\,\,Gpc^{-3}}$ \citep{Abbott2021}, though being on the lower side. Our minimum values are smaller compared to the value reported in \cite{Askar2017}. This supports the interpretation that the small value reported by our simulations is connected with the limitations imposed to the GC sub-population selected in our study.

The number of observable BBH merger rates in a 10 years span is quite small: our results would suggest that it would be unlikely to observe a BBH merger rate in both MW and M31 galaxies. In the future, results from the MOCCA-Survey Database II could be used to better constrain the BBH mass ratio and properties, taking advantages of the new stellar evolution features (as described in \citealt{Kamlah2021}) and the GC gas expulsion evolution included in the Monte Carlo method (as described in \citealt{Leveque2022}). 

The orbital properties of the BBHs that merged within 10 and 13 Gyr strongly depend on the formation channel of the binaries. Dynamical BBHs form due to close few-body encounters, and for this reason their orbits can have higher eccentricities and larger semi-major axis values compared to those BBHs that form through evolution of primordial binaries.

As shown in Fig. \ref{Fig:BHBHnumberDenistyDistribution}, the majority of the BBHs observable at 12 Gyr are located in the central galactocentric regions for both MW and M31. On the other hand, a constant mean number of BBHs per GC has been observed. This implies that the larger number of BBHs in the central region is correlated to the total number of GCs being more numerous in this region of the galactic halo, being the mean number of BBHs constant.

The radial distribution and the histograms at different galactocentric distances of semi-major axis, orbital eccentricity and mass ratio of observable BBHs shown in Fig. \ref{Fig:BHBHSurvivedPropMW} and Fig. \ref{Fig:BHBHSurvivedPropM31} could suggest stronger dynamical interactions involving BBHs in the central region of the galaxy halo. Indeed, the orbital eccentricity seems to be more thermal and the semi-major axis seems to be larger at smaller galactocentric distances (as shown in Fig. \ref{Fig:BHBHSurvivedRadialPropMW}). Instead, because of larger and less dense clusters in the outskirt of the galactic halo, the number of interactions that would thermalize the binaries could be expected to be smaller. This would imply a more circular orbital and smaller semi-major axis for BBHs hosted in GCs at larger galactocentric distances. This is supported by the galactocentric distance distribution of dynamical models in the galactic halo. As shown in Fig. \ref{Fig:dynamicalModelNDistribution}, the Standard models dominate the central region of the galactocentric halo. Given that the Standard models are more numerous compared to the BHS ones, it might be expected that the number of interactions that a BBHs undergo is larger in a Standard model than a BHS one.

We find around $1,000-3,000$ BHs are transported into the galactic NSC over a 12 Gyr time span. This might have interesting consequences on the overall population of BHs in a galactic nucleus. If NSCs form exclusively \textit{in-situ}, the fraction of stars turning into a BH can be retrieved by the initial mass function, and correspond to $\sim 8\times 10^{-4}$ for a \cite{Kroupa2001} mass function, thus a number of $N_{BH,in-situ} \sim 20,000$ for a MW-like NSC. If  a fraction $f$ of NSC mass is contributed by star cluster dispersal, the actual amount becomes $N_{BH,real} \sim (1-f)\,\, N_{BH,in-situ} + N_{BH,delivered}$. In our models, we assumed $f \sim 0.1$, leading to a negligible difference between full \textit{in-situ} formation and a ``mixed'' formation process. Combining our previous work and the present analysis, we can define a BH-transport efficiency as the ratio between the total number of delivered BHs and the total mass accreted into the galactic nucleus, this being around $\eta = 2,000 / (3.5\times 10^6) \sim 5.7 \times 10^{-4}\,\, M_\odot^{-1}$, and $N_{BH,delivered} = \eta \times M_{accreted}$. Thus, if a NSC is totally contributed by infalling clusters, we would expect a number of BHs $N_{BH,infall} = \eta \times M_{NSC} \sim 14,300$. This simplistic analysis provides us with a range of the total number of BHs that might be inhabiting the nuclear regions of MW and Andromeda, being this in the range $(1.4-2.2)\times 10^4$.

\subsection{Comparison of BH binary production in GCs and the field}
BHs are endpoints of the evolution of massive stars and they should be present in both dense stellar systems such as GCs, as well as in the galactic field. Comparing the expected number of BHs in such environments could be useful to understand the origin and the evolution of BHs and BH binaries as well. Using the results from the simulations carried out in this paper, we report the  total amount of BHs present at 12 Gyr for the MW population in Table \ref{Table:bh_binaries_population}. This includes the number of single BHs and the number of binaries containing at least one BH and different type of companions - main sequence (MS), stars outside the MS phase, white dwarf (WD), neutron stars (NS), and BH - are reported too. For this comparison, we did consider also the BHs and different types of BH binaries that would have been ejected from the GCs and hence would populate the Galactic halo - the numbers from MOCCA also include binaries that escaped their host cluster. These binaries were evolved from the time of escape up to 12 Gyr using the  {\tt StarTrack}  evolutionary code. For comparison, in the Table \ref{Table:bh_binaries_population} we provide the results  for Galactic field BH population \citep{Olejak2020}. The authors used {\tt StarTrack} population synthesis code \citep{Belczynski2008, Belczynski2020} to estimate current number of single BHs and BHs in binary systems that formed from the isolated stellar/binary evolution in three Milky Way components: bulge, disk and halo. They adopted individual star formation rate model and metallicity distribution for each component \citep[see Fig. 1 - 4 of][]{Olejak2020} and assumed isolated single/binary evolution of stars corresponding to total stellar mass of MW $M_{\rm MW} \approx 6.10 \times 10^{10} M_{\odot}$ \citep{Licquia2015}. The numbers provided for {\tt StarTrack} in Table \ref{Table:bh_binaries_population} correspond to total Galactic field population of bulge, disk and halo.

Our results for the number of BHs are of $3-4$ orders of magnitudes smaller compared to {\tt StarTrack} field population, and these differences can be explained by a smaller total stellar-mass in GCs used in our simulations. As already mentioned in Sec. \ref{sec:Method}, we constrained our initial models to a small fraction of the total GC initial mass function, and a region of the galactic potential, implying a total GCs mass population $\sim 75\%$ smaller than the whole population. Also, we excluded from our calculation the GCs that would have been dissolved during the Hubble time, that would compose of $\sim 10 \%$ of the total initial GC population mass in our simulations (and up to $\sim 25\%$ if we would consider an initial mass range between $10^3$ and $10^7\,M_\odot$). Hence, we would expect that numbers reported from the MOCCA simulations should increase at least by a factor of two, if we would consider in our simulations the BHs present in the very massive GCs in a  full mass-range GCIMF, and the BHs present in dissolved GCs. Indeed, it is expected that the most massive GCs would contribute importantly to the total number of BHs in GCs. Also, the number of BH binaries dynamically formed is expected to be important in such clusters. On the other hand, the contribution of old low mass GCs would be minimal, with small number of BHs formed in such system. Many BHs will escape such host clusters in its early stages of dynamical evolution and thus they are unlikely to form many BBHs \citep{Rastello2021,Torniamenti2022}. 

In Table \ref{Table:bh_binaries_population}, we also report the number of differed BBHs divided by the total initial stellar mass, which is $6.3\times 10^7\,\, M_\odot$ for the MOCCA simulations. When normalized by the total initial mass, our results are comparable (or even larger) to the number reported in {\tt StarTrack} simulations. The difference in the type of companion for the binaries containing a BH is related to the different stellar evolutionary formulae used in our simulations compared to the {\tt StarTrack} ones (for example, the outdated treatment for binary and stellar evolution in MOCCA-Survey Database I strongly underestimated the number of retained NSs in GC models, impacting the number of BH-NS binaries observed in our models), and by the dynamical interactions between BH and other objects inside the GCs. In fact, exchange in binary-single and binary-binary encounters is the most important process in the formation of binary systems with BHs. For example, dynamical encounters within GCs can lead to exchange interactions that can pair BHs with lower mass MS stars. Also, the presence of an IMBH in the GCs would influence the number of BH-MS and BH-WD binaries, preventing their formation, and it would reduce the number of merging BBHs in GCs \citep{Hong2020}. The dynamical interactions between binaries containing at least one BH and other objects in the GCs would strongly influence the fate of such binaries. In fact, according to Heggie-Hills law \citep{Heggie1975,Hills1975}, hard binaries (generally compact binaries) are more likely to get harder and soft binaries (generally wide binaries) to get softer due to the strong interaction between binary-single stars and binary-binary stars. Consequently, wide binaries (binaries whose stars would evolve or are BHs) would be dissolved, meanwhile hard binaries would get hard enough to survive the supernova (SN) events that their stars will undergo - and hence retrain the BHs that would be formed. Finally, single BHs can also form binaries during the interactions with other single BHs, this being more important in particular for BHS models.

As a result, when compared with the field production, the BH binaries efficiency (that is, the total number per unit of mass) in GCs is much larger than in the fields. In fact, as reported in Table \ref{Table:bh_binaries_population}, GCs are almost twice more efficient in producing binaries containing at least one BH, and even more efficient in producing BBHs compared to the field. Our results could be considered as a lower limit for the BH binaries efficiency. As mentioned before, a larger number of BHs and BH binaries are expected in clusters with masses larger than the one considered in this study. We would expect that the BH binaries efficiency could be larger of a factor of few, if we would consider such massive clusters. However, the nature of the dynamical formation of BH binaries would be imprinted in the orbital properties of the binaries, as shown in Fig. \ref{Fig:BHBHMergedHistrograms}. Finally, it must be cautioned that these results can be sensitive to the differences in binary/stellar evolution prescriptions used in both MOCCA and {\tt StarTrack}. In a future work, a more thorough comparison of the formation efficiency of BH binaries in the field and in GCs will be made using consistent stellar and binary evolution prescriptions between the two codes.

\begin{table*}
    \centering
    \begin{tabular}{cccccccc}
    \hline
    Model & $N_{BH,single}$ & $N_{BH,binaries}$ & $N_{BH-MS}$ & $N_{BH-Out\,of\,MS}$ & $N_{BH-WD}$ & $N_{BH-NS}$ & $N_{BH-BH}$\\
    \hline
    MOCCA & $8.3\cdot 10^4$ (94\%) & $6.4\cdot 10^3$ (98\%)  & 77 (0\%) & 3 (0\%)  & 42 (0\%) & 5 (100\%)    & $6.3\cdot 10^3$ (98\%)\\ 
   { \tt  StarTrack} (field) & $1.2 \cdot 10^8$ & $4.9 \cdot 10^6$ & $1.7 \cdot 10^5$  & $10^4$ & $1.4 \cdot 10^6$  &  $1.7 \cdot 10^5$  & $3.1 \cdot 10^6$\\
        \hline
    MOCCA & $ 10^{-3}$ & $ 10^{-4}$  & $ 10^{-6}$   (1\%) & $4\cdot 10^{-8}$  (0\%) & $7\cdot 10^{-7}$  (1\%) & $7\cdot 10^{-8}$  (0\%) & $10^{-4}$  (98\%) \\
     {\tt StarTrack} (field) & $2\cdot 10^{-3}$ & $8\cdot 10^{-5}$ & $3\cdot 10^{-6}$ (4\%) & $2\cdot 10^{-7}$  (0\%)& $2\cdot 10^{-5}$ (29\%) & $3\cdot 10^{-6}$ (4\%)& $5\cdot 10^{-5}$ (63\%)\\
    \hline
    \end{tabular}
    \caption{Total number of single BHs, number binaries containing at least one BH and the number of BH binaries with given companion are reported. In the top two rows, the actual number for MOCCA and  {\tt StarTrack} simulations are reported. Also, in parenthesis, the percentage of escaping binaries for the MOCCA simulation are reported. In the bottom two rows, the total numbers normalized by the total initial mass are reported. In parenthesis the percentage of each binary in respect of the total number of binaries containing a BH. The initial masses are of $6.3\times 10^7\,\, M_\odot$ for the MOCCA simulation and $6.3\times 10^{10}\,\, M_\odot$ for {\tt StarTrack} \citep{Olejak2020} respectively.} 
    \label{Table:bh_binaries_population}
\end{table*}
\section{Conclusion}
In this paper we expanded the study of the MW and M31 population simulated with the machinery introduced in Paper II, investigating the BH content of their GC population.

Summarizing our main results:
\begin{itemize}
    \item The kinematic properties of the MW population are in agreement with the observed ones - see Fig.  \ref{Fig:densityMapBajkowaPerEccMW}, Fig. \ref{Fig:densityMapBajkowaEccRadDistMW} and Fig. \ref{Fig:densityMapBajkowaRcMassDistMW}.
    The observed orbital properties distributions (that are, orbital eccentricity, pericenter distance, and the galactocentric distance for a circular orbit determined using the prescription in \cite{Cai2016}) in the MW population has been reproduced by our models, further confirming the reliability of our semi analytic procedure. This is a significant outcome for our machinery, since the GCs in external galaxy have been initially populated on a circular orbit, and subsequently they were modelled in elliptical orbits using the prescriptions in  \cite{Cai2016}.
    \item The mean GCs mass and the mean half-light radius for models that would harbour a BHS is larger than Standard and IMBH models (see Fig. \ref{Fig:dynamicalModelNDistribution}, Fig. \ref{Fig:dynamicalModelMeanMassDistribution} and Fig. \ref{Fig:dynamicalModelMeanRhDistribution}). On the other hand, Standard models are more numerous in the central region of the galaxy, with the number of IMBH and BHS comparable to the Standard ones at larger galactocentric distances.
    \item A maximum and a minimum value observable BBH merger rate has been determined, with value $\sim 20-35 \,\,\rm{yr^{-1}\,\,Gpc^{-3}}$ using the galaxy number density from the Illustris-1 simulation \citep{Vogelsberger2014}, and of $\sim 1.0-2.0 \,\,\rm{yr^{-1}\,\,Gpc^{-3}}$ using the extrapolated density of MWEGs from \cite{Abadie2010} respectively. The reported value are comparable to the value reported in previous works \citep{Mapelli2022,Banerjee2022,Askar2017}, although our results are on the lower-side with respect to other models (e.g. \citet{Rodriguez2018}). This could be due to the simplistic approach followed in calculating the merger rate. These differences can be explained by a smaller sub-sample of the whole GC population that have been considered and simulated in our machinery. 
    \item The signature of primordial or dynamically formed BBHs is imprinted in the orbital parameters for the merged binaries. Indeed, the dynamically formed binaries have greater mass ratios and more eccentric orbits than the primordial ones. Furthermore, it appears that the semi-major axis of the dynamically formed binaries is larger than that of the primordial binaries.  Conversely, primordial mergers are characterised by a nearly flat eccentricity distribution and a mass-ratio clearly peaked around 0.5. The primordial merger eccentricity distribution subtly implies that dynamics might have aided the merging process, owing to the fact that isolated stellar evolution generally predicts nearly circular BBH mergers.
    \item The observable BBHs at 12 Gyr that have been simulated in our models show different orbital properties for different galactocentric distances. The BBHs do show a larger (thermal) eccentricity, larger mass ratio ($>0.8$) and smaller semi-major axes ($<10^2 R_\odot$) at smaller galactocentric distances. These spatial evolution can be explained by denser GCs in the central region, enhancing the number of strong interactions between the BBHs and the other stars in the GCs.
    \item Most of the BH and BBH that are delivered to the NSC happens in the first $1-2$ Gyr of evolution, with a slow increase observed at later times. Also, the total number of BH binaries delivered to the NSC is $\sim 5\%$ of the total BH population delivered. A total of 1000-3000 BHs and 100-200 BBHs have transported into the nucleus over a time span of 12 Gyr. This implies a total number of  BHs and BBHs lurking in NSCs being of $N_{BHs} = (1.4-2.2)\times 10^4$ and $N_{BBHs} =  700-1,100$.
    \item The efficiency of BH binary formation, or the total number per unit of mass, can be significantly enhanced due to dynamics in GCs. In fact, compared to isolated stellar/binary evolution in the Galactic field, GCs are about twice as efficient at producing binaries with at least one BH and even more effective at generating BBHs.
\end{itemize}

In the future we would like to extend the study of the MW and M31 population to further investigate the super massive BH and nuclear star cluster masses build up \citep{Askar2022}. Also, we intend to simulate with our machinery other galaxies and the galaxies in the local Universe. We hope to restrict and identify the observational properties, evolutionary paths, and compact object content of GCs (such as IMBH, BHS, BBHs, and X-ray binaries), and also study the gravitational microlensing phenomena in GCs. Our simulation results might be utilized to calculate the BBH merger rate in the local Universe, as well as the event rates of TDEs between the SMBH and infalling GCs.

\section*{Acknowledgements}
We thank the anonymous reviewer for insightful comments that helped us clarify the presentation of the results in this paper. MG and AL were partially supported by the Polish National Science Center (NCN) through the grant UMO-2016/23/B/ST9/02732. AA acknowledges support from the Swedish Research Council through the grant 2017-04217. MAS acknowledges financial support from the European Union’s Horizon 2020 research and innovation programme under the Marie Skłodowska-Curie grant agreement No. 101025436 (project GRACE-BH, PI Manuel Arca Sedda). AO acknowledge support from the Polish National Science Center (NCN) grant Maestro (2018/30/A/ST9/00050).  AO is also supported by the Foundation for Polish Science (FNP) and a scholarship of the Minister of Education and Science (Poland).

\section*{Data Availability}
The data underlying this article will be shared on reasonable request to the corresponding author.
\bibliographystyle{mnras}
\bibliography{biblio}

\appendix
\section{Statistical testing of the studied populations  } \label[Appendix A]{Appendix}
To ensure that the obtained results are statistically consistent with the observed distributions and to check whether the GC parameters analyzed in this paper show dependence on the galactocentric distance, we used the Kolgomorov-Smirnov test (KS test).

The underlying continuous distributions of two separate samples are compared in the KS test. In particular, the ``two-sample'' KS test is used to determine if two samples come from the same distribution. The KS test  measures the distance between the two cumulative distribution functions (CDFs) of the two samples. For the ``two-sample'', the null hypothesis is that the two samples are drawn from the same distribution. For the KS tests carried out in this work, the null hypothesis is rejected for p < 0.05.

The KS test was first applied to compare the observed distributions of the pericenter distance, orbital eccentricity, and circular orbit distance for the MW GC population with the results from our simulations. The p-values for the ``two-sample'' KS tests are $0.06$ for the orbital eccentricity, $0.15$ for the  pericenter distance, and $0.05$ for the circular orbit distance distribution respectively. This comparison shows that our sampled models are marginally consistent with the observed distribution of orbital properties of MW clusters. The results refer to the data shown in Fig. \ref{Fig:densityMapBajkowaPerEccMW}, Fig. \ref{Fig:densityMapBajkowaEccRadDistMW}, and Fig. \ref{Fig:densityMapBajkowaRcMassDistMW}.

To verify that the GC property distributions obtained in our models show some dependence on the galactocentric distance we applied the KS test. We compared the galactocentric distance dependence of total number of GCs, mean GC mass, mean GC half-light radius, and total BH mass for different GC dynamic models (GC that includes a  BHS, IMBH or neither of those) with a uniform galactocentric distance distribution. For each set of simulation data, we carried out the ``two-sample''  KS test  against the null hypothesis that the cluster property has a uniform distribution in the galactocentric distance. In Tables \ref{Table:Ttest-results-uniform-MW} and \ref{Table:Ttest-results-uniform-M31}, we report the p-values from the ``two-sample'' KS tests for MW and M31 respectively. For MW, we find that the p-values returned from the ``two-sample'' KS tests  for galactocentric radius distributions of all properties for all types of cluster models are less than 0.05. This indicates that the galactocentric radius distributions of properties of these models are not consistent with a uniform distribution. Additionally, for distributions of the mean mass  for the Standard models and the mean half-light radius and total BH mass for the BHS models for M31 clusters, the obtained p-values were greater than 0.05.  For these cases, the galactocentric radius distribution of properties is consistent with a uniform distribution. For the other properties of M31 models, we found p-values lower than 0.05. This suggests that for these properties, our simulated distributions are not consistent with a uniform galactocentric distance distribution. The results reported in Table \ref{Table:Ttest-results-uniform-MW} refers to our results shown in the top row of Fig. \ref{Fig:dynamicalModelNDistribution}, Fig. \ref{Fig:dynamicalModelMeanMassDistribution}, Fig. \ref{Fig:dynamicalModelMeanRhDistribution}, and  Fig. \ref{Fig:dynamicalModelMeanMassBHDistribution}. Similarly, the  results reported in Table \ref{Table:Ttest-results-uniform-M31} refer to our results shown in the bottom row of the same Figures.

\begin{table*}
    \centering
    \begin{tabular}{ c@{\hskip 0.5in} c c c c}
    \hline
       Dynamical Model & Distribution of number of GCs & Mean mass distribution  & Mean half-light radius distribution & Total BH mass distribution\\
       \hline
        BHS & $5\times10^{-5}$  & $5\times10^{-5}$ & $6\times10^{-5}$  & $1\times10^{-11}$ \\
        IMBH & $5\times10^{-8}$  & $10^{-7}$  & $10^{-8}$  & $2\times10^{-11}$ \\ 
        Standard & $5\times10^{-19}$  & $5\times10^{-5}$ & $9\times10^{-6}$  & -  \\
    \hline
    \end{tabular}
    \caption{We carried out ``two-sample'' KS tests by comparing galactocentric distributions for the number, mean mass, mean half-light radius and total BH mass of simulated MW models with an assumed uniform in galactocentric radius distribution. For all these tests, the p-value are reported above. For the cases were the p-values are larger than 0.05, the null hypothesis cannot be rejected. The results from left to right refers to the data shown in the top row of Fig. \ref{Fig:dynamicalModelNDistribution}, Fig. 
 \ref{Fig:dynamicalModelMeanMassDistribution}, Fig. \ref{Fig:dynamicalModelMeanRhDistribution}, and  Fig. \ref{Fig:dynamicalModelMeanMassBHDistribution}, respectively.}
    \label{Table:Ttest-results-uniform-MW}
\end{table*}

\begin{table*}
    \centering
    \begin{tabular}{ c@{\hskip 0.5in} c c c c}
    \hline
       Dynamical Model & Distribution of number of GCs & Mean mass distribution  & Mean half-light radius distribution & Total BH mass distribution   \\
        \hline
        BHS & $2\times10^{-4}$  & $0.03$  & $0.09$  & $0.07$ \\
        IMBH & $2\times10^{-5}$  & $2\times10^{-4}$  & $2\times10^{-4}$  & $2\times10^{-4}$ \\ 
        Standard & $2\times10^{-5}$  & $0.07$  & $10^{-3}$  & -  \\
    \hline
    \end{tabular}
    \caption{We carried out ``two-sample'' KS tests by comparing galactocentric distributions for the number, mean mass, mean half-light radius and total BH mass of simulated M31 models with an assumed uniform in galactocentric radius distribution. For all these tests, the p-value are reported above. For the cases were the p-values are larger than 0.05, the null hypothesis cannot be rejected. The results from left to right refers to the data shown in the bottom row of Fig. \ref{Fig:dynamicalModelNDistribution}, Fig. \ref{Fig:dynamicalModelMeanMassDistribution}, Fig. \ref{Fig:dynamicalModelMeanRhDistribution}, and  Fig. \ref{Fig:dynamicalModelMeanMassBHDistribution}, respectively.}
    \label{Table:Ttest-results-uniform-M31}
\end{table*}

Furthermore, we have applied the KS test to verify that  the relative shape in dependence on galactocentric distance for different types of GC evolution scenario shown in Fig. \ref{Fig:dynamicalModelMeanRhDistribution} and  Fig. \ref{Fig:dynamicalModelMeanMassBHDistribution} are statistically meaningful for the mean mass distribution and the half-light radius distribution. For this study, we took as representative the distributions for the Standard model, and we compared it against the BHS and IMBH ones.  The p-values for the ``two-sample''  KS tests are reported in Table  \ref{Table:Ttest-results-standard}. For the mean mass distribution for both BHS and IMBH models in MW, and for the mean mass distribution for IMBH in M31 the distributions are comparable to the Standard models (p > 0.05 for these tests). Instead, statistically significant differences from the distribution of Standard model properties have been found for the distribution of the half-light radius for both BHS and IMBH models in both MW and M31, and for the mean mass distribution for BHS models in M31 respectively. This suggests that for the latter cases, the considered simulated distributions are not consistent with the Standard model distributions. 

\begin{table*}
    \centering
    \begin{tabular}{ c@{\hskip 0.5in} c c c c}
    \hline
    &  \multicolumn{2}{c}{MW} & \multicolumn{2}{c}{M31}\\
    \hline
        Dynamical Model &  Mean mass distribution & Mean half-light radius distribution & Mean mass distribution  & Mean half-light radius distribution  \\
        \hline
        BHS & $0.08$ & $0.01$ & $7\times10^{-7}$  & $7\times10^{-7}$ \\
        IMBH & $0.1$  & $4\times10^{-3}$  & $0.5$  & $7\times10^{-3}$ \\ 
    \hline
    \end{tabular}
    \caption{We carried out ``two-sample'' KS tests by comparing galactocentric distributions for the mean mass and mean half-light radius of the BHS and IMBH models with the same distributions for the Standard models for the simulated population of GCs in MW (first two columns) and M31 (last two columns). For all these tests, the p-value are reported above. For the cases were the p-values are larger than 0.05, the null hypothesis cannot be rejected. The results refer to the data shown in Fig. \ref{Fig:dynamicalModelMeanMassDistribution} and Fig. \ref{Fig:dynamicalModelMeanRhDistribution}. }
    \label{Table:Ttest-results-standard}
\end{table*}

Finally, we applied a KS test for the spatial distribution of the mean values in radial bins for the semi-major axis, eccentricity and mass ratio for non-merging BBHs, comparing them with a uniform distribution  in galactocentric distances. The p-values for the ``two-sample'' KS test are  $5\times10^{-5}$ for the orbital eccentricity,  $10^{-9}$ for the mass ratio, and $10^{-5}$ for the semi-major axis distribution, respectively. As it is possible to note, there are statistical differences for all properties. This suggests that for these cases, our simulated distributions are not consistent with a uniform galactocentric distance distribution. These results refer to the data shown in Fig. \ref{Fig:BHBHSurvivedRadialPropMW}, from left to right respectively. 

\bsp
\label{lastpage}

\end{document}